\documentclass[10pt]{iopart}

    \usepackage{iopams}
    \usepackage{amssymb}
    \usepackage{amsfonts}
    \usepackage{amstext}
    \usepackage{graphicx}
    \usepackage{setstack}
    \usepackage{amscd}
    \usepackage{color}
    \usepackage{ulem}

\newcommand{\ihbar}{\imath \hbar}
\newcommand{\Te}{\mathbb{T}e}

\newcommand{\dist}{\mathrm{dist}_{\mathrm{FS}}}
\newcommand{\Ran}{\mathrm{Ran}}

\begin{document}

\title[Almost quantum adiabatic dynamics and generalized TDWO]{Almost quantum adiabatic dynamics and generalized time dependent wave operators}

\author{David Viennot}
\address{Institut UTINAM (CNRS UMR 6213, Universit\'e de Franche-Comt\'e), 41bis Avenue de l'Observatoire, BP1615, 25010 Besan\c con cedex, France.}

\begin{abstract}
We consider quantum dynamics for which the strict adiabatic approximation fails but which do not escape too far from the adiabatic limit. To treat these systems we introduce a generalisation of the time dependent wave operator theory which is usually used to treat dynamics which do not escape too far from an initial subspace called the active space. Our generalisation is based on a time dependent adiabatic deformation of the active space. The geometric phases associated with the almost adiabatic representation are also derived. We use this formalism to study the adiabaticity of a dynamics surrounding an exceptional point of a non-hermitian hamiltonian. We show that the generalized time dependent wave operator can be used to correct easily the adiabatic approximation which is very unperfect in this situation.
\end{abstract}

\section{Introduction}
The numerical study of complex quantum dynamical systems, as the interaction of a molecule with a strong laser field, leads to need for a long computational times and large computer memory capacity when use is made of a wave packet approach, which involves a direct integration of the time-dependent Schr\"odinger equation. Moreover the theoretical study of such systems is difficult because the time-dependent wave function involves components belonging to the whole Hilbert space. It is then interesting to approach the true dynamics by an effective dynamics within a small subspace, called an active space. Moreover it is important to be able to compare the true and the effective dynamics.\\
The time-dependent wave operator theory \cite{killingbeck,jolicard} can be used if the dynamics does not escape too far from an inital small subspace (the meaning of ``espace too far from a subspace'' will be precisely defined in the next section). The effective dynamics within the active space, which approaches the true dynamics, is governed by an effective Hamiltonian $H^{eff} = P_0H\Omega$ (where $H$ is the true Hamiltonian, $P_0$ is the projector onto the active space and $\Omega$ is the time-dependent wave operator). The time-dependent wave operator is a comparison of the true and the effective dynamics and can be used to deduce the true dynamics from the effective dynamics. The time-dependent wave operators are a generalization of the M\o ller wave operators $\Omega^{\pm} = \lim_{t \to \mp \infty} e^{-\ihbar^{-1} Ht}e^{\ihbar^{-1} H_0t}$ which compares the scattering dynamics induced by a true Hamiltonian $H$ with the scattering dynamics induced by a simpler Hamiltonian $H_0$.\\
The main assumption of the time-dependent wave operator theory -- the dynamics does not escape too far from a fixed subspace -- can be a strong limitation. When it is not applicable, the adiabatic approximation \cite{messiah,nenciu,nenciu2,joye} can be used for some systems. The main principle of the adiabatic approximation is that the dynamics remains in the neighbourhood of a small time-dependent subspace generated by some instantaneous eigenvectors. The approximate dynamics is then governed by the effective Hamiltonian $H^{eff} = P_0 H P_0 - \ihbar \dot P_0 P_0$ (where $P_0$ is the orthogonal projector onto the adiabatic active space spaned by the few instantaneous eigenvectors, and the dot denotes the time derivative). $\ihbar \dot P_0 P_0$ is associated with the geometric (Berry) phase \cite{berry,viennot1}.\\
The conditions of the adiabatic approximation can be very restrictive (the Hamiltonian variations must be slow and a gap condition between the eigenvalues is required). In this paper we present a generalization of the active space method useful for dynamics which does not escape too far from an adiabatic subspace. In contrast with the strict adiabatic approximation, our generalization is not an asymptotic limit but an approach similar to the time-dependent wave operator theory. A generalization of the time dependent wave operators is indeed used to compare the dynamics within the active space and the true dynamics. This approach is then a mixing between the two previous approaches, as it is shown by the structure of the effective Hamiltonian of this almost adiabaticity, $H^{eff} = P_0H\Omega - \ihbar \dot P_0 \Omega$.\\
Next section recalls the main properties of the time-dependent wave operator theory. Section III introduces the almost adiabaticity and the associated generalized time-dependent wave operators. Section IV studies the geometric phases associated with the almost adiabatic representation. Section V studies the problem of the adiabatic approximation for a dynamics surrounding an exceptional point of a non-hermitian hamiltonian. The use of the almost adiabatic formalism helps to increase the accuracy of the description with respect to a strict adiabatic approximation. A simple analytical two level system is treated but also the case of the molecule $H_2^+$. The appendix presents the demonstration of the equation satisfied by the generalized time-dependent wave operator. It is interesting to note that this demonstration borrows ideas from the demonstration of the equation satisfied by the usual time-dependent wave operator and from the demonstrations of the adiabatic theorems.

\section{Space of projectors and time-dependent wave operators}
Let $G_m(\mathcal H) = \{P \in \mathcal B(\mathcal H), P^2=P, P^\dagger=P, \tr P = m \}$ be the space of rank $m$ orthogonal projectors of the Hilbert space $\mathcal H$ ($\mathcal B(\mathcal H)$ denotes the set of bounded operators of $\mathcal H$). If $\mathcal H$ is finite dimensional, i.e. $\mathcal H \simeq \mathbb C^n$, $G_m(\mathbb C^n)$ is a complex manifold called a complex grassmanian \cite{rohlin}. This manifold is endowed with a K\"ahlerian structure \cite{nakahara}, and particularly with a distance (called the Fubini-Study distance) defined by
\begin{equation}
\forall P_1,P_2 \in G_m(\mathbb C^n), \quad \dist(P_1,P_2) = \arccos|\det Z_1^\dagger Z_2|^2
\end{equation}
where $Z_1, Z_2 \in \mathfrak M_{n \times m}(\mathbb C)$ are the matrices of two arbitrary orthonormal basis of $\Ran P_1$ and $\Ran P_2$ expressed in an orthonormal basis of $\mathbb C^n$ ($\Ran P$ denotes the range of $P$). We can note that $0 \leq \dist(P_1,P_2) \leq \frac{\pi}{2}$. The Fubini-Study distance measures the ``quantum compatibility'' between the two subspaces $\Ran P_1$ and $\Ran P_2$ in the sense that $\dist(P_1,P_2) = \frac{\pi}{2}$ if and only if $\Ran P_1^\bot \cap \Ran P_2 \not=\{0\}$ or $\Ran P_1 \cap \Ran P_2^\bot \not=\{0\}$, i.e. there exists a state of $\Ran P_1$ for which the probability of obtaining the same measures as that with a system in a state of $\Ran P_2$ is zero \cite{viennot1}. For infinite dimensional Hilbert space, it is possible to define a manifold $G_m(\mathbb C^\infty)$ endowed with a K\"ahlerian structure by using the inductive limit techniques \cite{rohlin}.\\
Let $P_0,P \in G_m(\mathcal H)$ be such that $\dist(P_0,P) < \frac{\pi}{2}$. We call wave operator associated with $\Ran P_0$ and $\Ran P$ the operator $\Omega$ defined by
\begin{equation}
\Omega = P(P_0PP_0)^{-1}
\end{equation}
where $(P_0PP_0)^{-1} = P_0(P_0PP_0)^{-1} P_0$ is the inverse of $P$ within $\Ran P_0$ (it exists only if $P$ is not too far from $P_0$, i.e. $\dist(P,P_0) < \frac{\pi}{2}$). Usually the wave operators are used to solve eigenequations \cite{killingbeck}. In that case, we solve an effective eigenequation $H^{eff} \psi_0 = \lambda \psi_0$ where $H^{eff} = P_0 H \Omega \in \mathcal L(\Ran P_0)$ is the effective Hamiltonian within $\Ran P_0$ ($H \in \mathcal B(\mathcal H)$ is the true self-adjoint Hamiltonian). We recover the true eigenvector associated with $\lambda$, $H\psi = \lambda \psi$, by $\psi = \Omega \psi_0 \in \Ran P$ ($\psi_0 = P_0 \psi$). $\Omega$ is called Bloch wave operator and is obtained by solving the Bloch equation
\begin{equation}
H \Omega = \Omega H \Omega \iff [H,\Omega] \Omega = 0
\end{equation}
Since $\Omega^2 = \Omega$, the Bloch wave operator can be viewed as a non-linear generalization of an eigenprojector (an eigenprojector satisfying $[H,P] = 0$ with $P^2=P$). Physically, the Bloch wave operator compares the approximate eigenstates within $\Ran P_0$ (which is called the active subspace) with the associated true eigenstates. The Bloch equation can be numerically solved by the RDWA method (\textit{Recursive Distorded Wave Approximation}) \cite{jolicard} and a relevant active space $\Ran P_0$ can be numerically selected by using the WOSA method (\textit{Wave Operator Sorting Algorithm}) \cite{wyatt}.\\
In a same manner, in order to compare an approximate quantum dynamics within an active space $\Ran P_0$ with the true dynamics, we can introduce the time-dependent wave operator \cite{jolicard} :
\begin{equation}
\Omega(t) = P(t)(P_0 P(t) P_0)^{-1}
\end{equation}
where $(P_0 P(t) P_0)^{-1}$ is still the inverse within $\Ran P_0$, and where $t \mapsto P(t) \in G_m(\mathcal H)$ is the solution of the Schr\"odinger-von Neumann equation :
\begin{equation}
\label{SvN}
\ihbar \dot P(t) = [H(t),P(t)] \qquad P(0) = P_0
\end{equation}
$H(t) \in \mathcal B(\mathcal H)$ being the self-adjoint time-dependent Hamiltonian. We can then solve the effective Schr\"odinger equation within $\Ran P_0$, $\ihbar \partial_t \psi_0(t) = H^{eff}(t)\psi_0(t)$, where $H^{eff}(t) = P_0 H(t) \Omega(t) \in \mathcal L(\Ran P_0)$ is the effective Hamiltonian, and we recover the true wave function, $\ihbar \partial_t \psi(t) = H(t) \psi(t)$, by $\psi(t) = \Omega(t)\psi_0(t)$ ($P_0 \psi(t) = \psi_0(t)$). The time-dependent wave operator can be usued only if the dynamics does not escape too far from the initial subspace, i.e. $\forall t$, $\dist(P(t),P_0) < \frac{\pi}{2}$. Since $P(t) = U(t,0) P_0 U(t,0)^\dagger$, where $U(t,0) \in \mathcal U(\mathcal H)$ is the evolution operator ($\ihbar \dot U(t,0) = H(t)U(t,0)$, $U(0,0)= 1$; $\mathcal U(\mathcal H)$ denotes the set of unitary operators of $\mathcal H$), we can also write
\begin{equation}
\Omega(t) = U(t,0)(P_0U(t,0)P_0)^{-1}
\end{equation}
By using this expression, it is not difficult to prove that the time-dependent wave operator satisfies
\begin{equation}
\ihbar \dot \Omega(t) = H(t) \Omega(t) - \Omega(t)H(t)\Omega(t) = [H(t),\Omega(t)]\Omega(t) \quad \Omega(0)=P_0
\end{equation}
Usually this last equation is written
\begin{equation}
(H(t)-\ihbar \partial_t) \Omega(t) = \Omega(t) (H(t)-\ihbar \partial_t) \Omega(t)
\end{equation}
which is right since $\Omega(t)\dot \Omega(t) = 0$. The time-dependent wave operator satisfies then a Bloch equation with the Floquet Hamiltonian $H_F(t) = H(t) - \ihbar \partial_t$ in the extended Hilbert space $\mathcal H \otimes L^2_0([0,T],dt)$ ($T$ being the duration of the dynamics and $L^2_0([0,T],dt)$ denotes the space of square integrable functions of $[0,T]$ with periodic limit conditions). We can then apply the numerical methods solving the Bloch equation to compute the time-dependent wave operator (see ref. \cite{jolicard} to have a complete presentation of the use of the generalized Floquet theory with time-dependent wave operators).

\section{Almost adiabaticity and generalized time-dependent wave operators}
Let $t \mapsto H(t) \in \mathcal B(\mathcal H)$ be a self-adjoint time-dependent Hamiltonian. Let $t\mapsto P_0(t) \in G_m(\mathcal H)$ be a $\mathcal C^2$ instantaneous eigenprojector:
\begin{equation} 
\forall t, \quad [H(t),P_0(t)]=0
\end{equation}
Let $U_T(s,0) \in \mathcal U(\mathcal H)$ be the evolution operator for the reduce time $s=t/T$, $\ihbar T \partial_s U_T(s,0) = H(sT) U_T(s,0)$ with $U_T(0,0)=1$. The dynamics of the quantum system is said adiabatic if
\begin{equation}
\forall s, \quad U_T(s,0)P_0(0) = P_0(sT) U_T(s,0) + \mathcal O(1/T)
\end{equation}
for $T$ in the neighbourhood of $+ \infty$. The adiabaticity is realized if $H(t)$ and $P_0(t)$ satisfy an adiabatic theorem \cite{messiah,nenciu,nenciu2,joye}. We can reformulate the adiabatic assumption. Let $t \mapsto P(t) \in G_m(\mathcal H)$ be the solution of the Schr\"odinger-von Neumann equation
\begin{equation}
\ihbar \dot P = [H(t),P(t)] \qquad P(0)=P_0(0)
\end{equation}
The dynamics is adiabatic if
\begin{equation}
\forall t, \quad \dist(P(t),P_0(t)) = \mathcal O(1/T)
\end{equation}

The adiabatic assumption is realized if the following Riemann-Lebesgue like lemma is satisfied \cite{messiah}: $\forall \psi_a \in \Ran P_0(t)$
\begin{eqnarray}
& & \sum_{\psi_b \in \Ran P_0(t)^\bot} \left| \int_0^s e^{\ihbar^{-1} T \int_0^{s'} \Delta E_{ab}(s'')ds''} \frac{\langle \psi_a(s')|\dot H(s')|\psi_b(s')\rangle}{\Delta E_{ab}(s')} ds' \right|^2 \nonumber \\ & & \qquad \qquad = \mathcal O(1/T)
\end{eqnarray}
where $\{\psi_a\}_a$ is a $\mathcal C^1$ eigenbasis of $\mathcal H$, and $\Delta E_{ab}(s) = E_b(s)-E_a(s)$, $E_a$ being the eigenvalue associated with $\psi_a$. In practice this needs a gap condition ($\forall t$ the eigenvalues associated with $\Ran P_0(t)$ must remain relatively far from the other eigenvalues) and slow variations of the Hamiltonian (in order to be close to the idealization $T \to + \infty$). This can be very drastic.\\
We consider now systems for which these conditions are not applicable (too fast variations or spectrum with too small gaps), but which satisfy the following weakest adiabatic assumption
\begin{equation}
\exists r < \frac{\pi}{2}, \quad \forall t, \quad \dist(P(t),P_0(t)) \leq r
\end{equation}
We said then that the quantum system is almost adiabatic. $r$ is called the almost adiabaticity radius and the strict adiabaticity corresponds to $r \to 0$. In contrast with the strict adiabaticity where the dynamics must not significantly escape from the adiabatic subspace $\Ran P_0(t)$, the almost adiabaticity requires only that the dynamics does not escape too far from the adiabatic subspace ($\dist(P(t),P_0(t))< \frac{\pi}{2}$). Since $\Ran P(t)$ is close to the adiabatic subspace $\Ran P_0(t)$ we call it the limbo space.\\

In the almost adiabatic case, the dynamics within the adiabatic subspace $\Ran P_0(t)$ is a wrong approximation of the true dynamics, it is then necessary to compare the quantum dynamics within the adiabatic space with the true dynamics involving limbo states. This is achieved with the following generalization of the time-dependent wave operator:
\begin{equation}
\Omega(t) = P(t)(P_0(t)P(t)P_0(t))^{-1}
\end{equation}
$(P_0(t)P(t)P_0(t))^{-1}$ is the inverse within $\Ran P_0(t)$ (it exists because of the almost adiabatic assumption). The essential difference with the usual time-dependent wave operator is that $P_0$ is time-dependent, nevertheless the generalized time-dependent wave operator satisfies the equation
\begin{equation}
\label{eqWO}
\ihbar \dot \Omega(t) = H(t)\Omega(t) - \Omega(t)H(t)\Omega(t) + \ihbar \Omega(t)\dot \Omega(t)
\end{equation}
which can be rewritten as a Bloch equation
\begin{equation}
(H(t)-\ihbar \partial_t) \Omega(t) = \Omega(t) (H(t)-\ihbar \partial_t) \Omega(t)
\end{equation} 
This property is proved in \ref{appA}. It can be surprising that the generalized time-dependent wave operator obeys to the same equation that the usual time-dependent wave operator, but we can note that in the usual case we consider only the solutions of the Bloch equation such that $\Omega \dot \Omega=0$, whereas the solutions of the generalized case must be such that $\Omega \dot \Omega \not=0$. An other important difference concerns the properties of $\Omega$ in the Floquet theory. In the two cases we have in $\mathcal H$, $P_0 \Omega = P_0$ with $P_0$ time-dependent or not. But in the extended Hilbert space $\mathcal H \otimes L^2_0([0,T],dt)$, for the usual time-dependent wave operator, we have
\begin{equation}
\hat P_0 = P_0 \otimes 1 \Rightarrow \hat P_0 \hat \Omega = \hat P_0
\end{equation}
where $\hat .$ denotes the representation of a time-dependent operator of $\mathcal H$ as an operator of $\mathcal H \otimes L^2_0([0,T],dt)$. For the generalized wave operator, we have
\begin{equation}
P_0(t) = \sum_{n \in \mathbb Z} P_{0n} e^{\imath n \omega t}
\end{equation}
where $\omega = \frac{2 \pi}{T}$ is the frequency of the artificial period of the generalized Floquet theory, and $P_{0n} \in \mathcal B(\mathcal H)$ are the Fourier components of $P_0$. We can note that
\begin{equation}
\label{convolution}
P_0(t)^2 = P_0(t) \iff \forall n, \sum_{q \in \mathbb Z} P_{0n-q} P_{0q} = P_{0n}
\end{equation}
Let $\{|n \rangle = e^{\imath n \omega t} \}_{n \in \mathbb Z}$ be the canonical basis of $L^2_0([0,T],dt)$. The components of $P_0(t)$ with respect to this basis are
\begin{equation}
\langle m|P_0|q \rangle_{L^2_0([0,T],dt)} = \int_0^T e^{- \imath m \omega t} P_0(t) e^{\imath q \omega t} dt = P_{0q-m}
\end{equation}
and then
\begin{equation}
\hat P_0 = \sum_{q,m \in \mathbb Z} P_{0q-m} \otimes |m \rangle \langle q|
\end{equation}
and finally
\begin{equation}
\hat P_0^2 = \sum_{ml \in \mathbb Z} \left( \sum_{q \in \mathbb Z} P_{0q-m}P_{0l-q} \right) \otimes |m \rangle \langle l| \not= \hat P_0
\end{equation}
$\hat P_0$ is then not a projector in $\mathcal H \otimes L^2_0([0,T],dt)$ and then $\hat P_0 \hat \Omega \not= \hat P_0$ (this is not surprising since the Floquet representation consists of Fourier transforms which transform the products into convolutions, then $P_0(t)^2 = P_0(t) \Rightarrow \hat P_0 \ast \hat P_0 = \hat P_0$ this is precisely the equation \ref{convolution}).\\

$H^{eff}= (P_0 H - \ihbar \dot P_0) \Omega$ governs the effective dynamics since (see \ref{appA5})
\begin{equation}
\label{decomp}
U(t,0)P_0(0) = \Omega(t) U^{eff}(t,0)
\end{equation}
where the effective evolution operator $U^{eff}(t,0) \in \mathcal L(\Ran P_0(0),\Ran P_0(t))$ ($U^{eff}$ is not unitary, it is just invertible) is the solution of
\begin{equation}
\ihbar \dot U^{eff}(t,0) = H^{eff}(t) U^{eff}(t,0) \qquad U^{eff}(0,0) = P_0(0)
\end{equation}

\section{Geometric phases in almost adiabaticity}
\subsection{One dimensional case}
We consider the case where $\dim \Ran P_0 = 1$. Let $\phi_0(t) \in \mathcal H$ be the considered normalized eigenvector:
\begin{equation}
H(t)\phi_0(t) = \lambda(t)\phi_0(t), \qquad P_0(t) = |\phi_0(t) \rangle \langle \phi_0(t)|
\end{equation}

By application of the equation \ref{decomp}, the true wave function $\psi(t) \in \mathcal H$ solution of the Schr\"odinger equation with the inital condition $\psi(0)=\phi_0(0)$ can be written in the almost adiabatic representation as
\begin{equation}
\psi(t) = c(t) \Omega(t) \phi_0(t) \qquad c(t) \in \mathbb C^*
\end{equation}
where the generalized time-dependent wave operator is used to transform the wrong adiabatic approximation of the wave function $c(t)\phi_0(t)$ into the true wave function. By inserting this expression of $\psi$ in the Schr\" odinger equation, we find
\begin{eqnarray}
& & \dot c(t) \Omega(t) \phi_0(t) + c(t)\dot \Omega(t)\phi_0(t)+c(t)\Omega(t)\frac{d\phi_0(t)}{dt} \nonumber \\
& & = -\ihbar^{-1} c(t)H(t)\Omega(t)\phi_0(t)
\end{eqnarray}
We project this equation onto $\langle \phi_0(t)|\Omega^{-1}(t)$ where $\Omega^{-1}(t) = P_0(t)P(t)$ is the pseudo-inverse of $\Omega$ ($\Omega^{-1} \Omega = P_0$).
\begin{eqnarray}
\dot c(t) & = & \left(-\ihbar^{-1} \langle \phi_0(t)|\Omega^{-1}(t)H(t)\Omega(t)|\phi_0(t)\rangle - \langle \phi_0(t)|\partial_t|\phi_0(t)\rangle \nonumber \right. \\
& & \qquad \left. - \langle \phi_0(t)|\Omega^{-1}(t)\dot \Omega(t)|\phi_0(t) \rangle \right)c(t)
\end{eqnarray}
Finally we have
\begin{equation}
\label{adiabtransp}
\psi(t) = e^{-\ihbar^{-1} \int_0^t \lambda^{eff}(t')dt' - \int_0^t A(t')dt' - \int_0^t \hat \eta(t')dt'} \Omega(t) \phi_0(t)
\end{equation}
with
\begin{eqnarray}
\lambda^{eff}(t) & = & \langle \phi_0(t)|\Omega^{-1}(t)H(t)\Omega(t)|\phi_0(t)\rangle \\
A(t) & = & \langle \phi_0(t)|\partial_t|\phi_0(t)\rangle \\
\hat \eta(t) & = & \langle \phi_0(t)|\Omega^{-1}(t) \dot \Omega(t)|\phi_0(t)\rangle
\end{eqnarray}
We see that the correct adiabatic approximation in almost adiabaticity $e^{-\ihbar^{-1} \int_0^t \lambda^{eff}(t')dt' - \int_0^t A(t')dt' - \int_0^t \hat \eta(t')dt'}\phi_0(t)$ differs from the usual adiabatic approximation by two points:
\begin{itemize}
\item the presence of an extra geometric phase $e^{- \int_0^t \hat \eta(t')dt'}$. Such a wave operator geometric phase has been already observed in the different context of the generalization of the Stokes theorem to geometric phases associated with a resonances crossing \cite{viennot2}.
\item the replacement of the usual dynamical phase $e^{- \ihbar^{-1} \int_0^t \lambda(t')dt'}$ by an effective dynamical phase $e^{- \ihbar^{-1} \int_0^t \lambda^{eff}(t')dt'}$. 
\end{itemize}
By setting $\tilde \psi(t) = \Omega(t) \psi_0(t)$, equation (\ref{adiabtransp}) becomes
\begin{equation}
\psi(t) = e^{-\ihbar^{-1} \int_0^t \frac{\langle \tilde \psi(t')|H(t')|\tilde \psi(t')\rangle}{\langle \tilde \psi(t')|\tilde \psi(t') \rangle} dt' - \int_0^t \frac{\langle \tilde \psi(t')|\partial_{t'}|\tilde \psi(t')\rangle}{\langle \tilde \psi(t')|\tilde \psi(t') \rangle} dt'} \tilde \psi(t)
\end{equation}
This is precisely the expression of the (Aharonov-Anandan) geometric phase for a cyclic non-adiabatic evolution \cite{aharonov}. Nevertheless, in this case the cyclicity cannot be completely assumed. We can assume that $P_0(T) = P_0(0)$ but we have only $\dist(P(T),P(0)) < \frac{\pi}{2}$ (the evolution is ``almost cyclic'').\\
By using equation (\ref{eqWO}) we have $\ihbar \Omega^{-1}(t) \dot \Omega(t) = \Omega^{-1}(t)H(t)\Omega(t) - P_0(t)H(t)\Omega(t) + \ihbar P_0(t) \dot \Omega(t)$, and then $\hat \eta(t) = -\ihbar^{-1}(\lambda^{eff}(t) - \lambda(t)) + \langle \phi_0(t)|\dot \Omega(t)|\phi_0(t)\rangle$. Equation (\ref{adiabtransp}) can be then rewritten with the usual dynamical phase as follows:
\begin{equation}
\psi(t) = e^{-\ihbar^{-1} \int_0^t \lambda(t')dt' - \int_0^t A(t')dt' - \int_0^t \eta(t')dt'} \Omega(t) \phi_0(t)
\end{equation}
with a reduced wave operator geometric phase generated by
\begin{equation}
\eta(t) = \langle \phi_0(t)|\dot \Omega(t)|\phi_0(t) \rangle
\end{equation}

\subsection{Multidimensional case}
Now we consider that $\dim \Ran P_0 = m >1$. Let $\hat H^{eff}(t) = P_0(t)H(t)\Omega(t) \in \mathcal L(\Ran P_0(t))$ be an effective Hamiltonian within $\Ran P_0$. Even if $\Ran P_0$ is generated by a basis of eigenvectors of $H$, these vectors are not eigenvectors of $\hat H^{eff}$. We consider then the effective eigenvectors $\phi^{eff}_{0a} \in \Ran P_0$ ($a=1,...,m$):
\begin{equation}
\hat H^{eff}(t) \phi^{eff}_{0a}(t) = \lambda^{eff}_{0a}(t) \phi^{eff}_{0a}(t)
\end{equation}
We can note that by construction $\lambda_{0a}^{eff}(t)$ is an eigenvalue of $P_0(t)H(t)$ with the associated eigenvector $\Omega(t)\phi^{eff}_{0a}(t)$. The true wave function which is the solution of the Schr\"odinger equation with the initial condition $\psi(0) = \phi^{eff}_{0a}(0) = \phi_{0a}(0)$ ($\phi_{0a}$ is an eigenvector of $H$) is in the almost adiabatic representation (by application of equation \ref{decomp}):
\begin{equation}
\label{step1}
\psi(t) = \sum_{b=1}^m c_b(t) \Omega(t) \phi^{eff}_{0b}(t)
\end{equation}
Since $\hat H^{eff}$ is not self-adjoint, the effective eigenvectors do not form an orthonormal basis. Let $T(t) \in \mathfrak M_{m \times m}(\mathbb C)$ be the matrix defined by
\begin{equation}
T_{bc}(t) = \langle \phi^{eff}_{0b}(t)|\phi^{eff}_{0c}(t) \rangle
\end{equation}
we define then
\begin{equation}
\langle \phi^{eff}_{0b}(t)*| = \sum_{c=1}^m [T^{-1}(t)]_{bc} \langle \phi^{eff}_{0c}(t)|
\end{equation}
We have then
\begin{equation}
\langle \phi^{eff}_{0b}(t)*|\phi^{eff}_{0c}(t) \rangle = \delta_{bc}
\end{equation}
By projecting on $\langle \phi^{eff}_{0c}(t)*|$ the equation obtained by inserting the expression \ref{step1} in the Schr\"odinger equation, we obtain
\begin{eqnarray}
\psi(t) & = & \sum_{b=1}^m \left[\Te^{- \ihbar^{-1} \int_0^t E^{eff}(t')dt' - \int_0^t A^{eff}(t')dt' - \int_0^t \eta(t')dt'}\right]_{ba} \nonumber \\
& & \qquad \qquad \times \Omega(t) \phi^{eff}_{0b}(t)
\label{adiabtranspNA}
\end{eqnarray}
where $\Te$ is the time ordered exponential (i.e. a Dyson series) and where the matrices $E^{eff},A^{eff},\eta \in \mathfrak M_{m\times m}(\mathbb C)$ are defined by
\begin{equation}
E^{eff}(t)  =  \left(\begin{array}{ccc} \lambda^{eff}_1(t) & & 0 \\ & \ddots & \\ 0 & & \lambda^{eff}_m(t) \end{array} \right) 
\end{equation}
\begin{equation}
A^{eff}(t)  =  \left(\begin{array}{ccc} \langle \phi^{eff}_{01}(t)*|\partial_t|\phi^{eff}_{01}(t) \rangle & ... & \langle \phi^{eff}_{01}(t)*|\partial_t|\phi^{eff}_{0m}(t) \rangle \\ \vdots & \ddots & \vdots \\ \langle \phi^{eff}_{0m}(t)*|\partial_t|\phi^{eff}_{01}(t) \rangle & ... & \langle  \phi^{eff}_{0m}(t)*|\partial_t|\phi^{eff}_{0m}(t) \rangle \end{array} \right) 
\end{equation}
\begin{equation}
\eta  =  \left(\begin{array}{ccc} \langle \phi^{eff}_{01}(t)*|\dot\Omega(t)|\phi^{eff}_{01}(t) \rangle & ... & \langle \phi^{eff}_{01}(t)*|\dot\Omega(t)|\phi^{eff}_{0m}(t) \rangle \\ \vdots & \ddots & \vdots \\ \langle  \phi^{eff}_{0m}(t)*|\dot\Omega(t)|\phi^{eff}_{01}(t) \rangle & ... & \langle \phi^{eff}_{0m}(t)*|\dot\Omega(t)|\phi^{eff}_{0m}(t) \rangle \end{array} \right)
\end{equation}

As for the one dimensional case, the ``non-abelian phase'' is generated by an effective eigenvalue matrix in place of the true eigenvalue matrix and by an extra geometric phase generator associated with the wave operator. We can note that $A^{eff}$ is not physically different from of the usual geometric phase generator $A$. Let $M(t) \in \mathcal{GL}(\Ran P_0(t))$ be the passage matrix tranforming the basis of $\Ran P_0(t)$ constituted by the eigenvectors of $H(t)$ into the basis $\{\phi^{eff}_{0a}\}_a$. We have
\begin{equation}
A^{eff}(t) = M^{-1}(t)A(t)M(t) + M^{-1}(t)\dot M(t)
\end{equation}
The transformation of $A$ to $A^{eff}$ is then just a gauge change, $A^{eff}$ and $A$ have then the same physical meaning.\\
Remark: in contrast with $M$, $\Omega$ cannot be viewed as a gauge change since it is not invertible.

\subsection{Computation of the wave function in almost adiabaticity}
Formulae (\ref{adiabtransp}) and (\ref{adiabtranspNA}) have not interest if we cannot compute easily the generalized time-dependent wave operator $\Omega(t)$. We have shown that it is solution of a Bloch equation in the extended Hilbert space of the Floquet theory. Unfortunately we cannot use the RDWA technique to integrate the equation because the time-dependent projector $P_0(t)$ is not a projector in the extended Hilbert space.\\
Let $X(t) = Q_0(t)\Omega(t)P_0(t)$ be the reduced wave operator ($\Omega(t) = P_0(t) + X(t)$), where $Q_0(t) = 1- P_0(t)$ (the strict adiabaticity corresponds to $X(t) \simeq 0$). Let $K(t) = \ihbar (\dot P_0(t) P_0(t) + \dot Q_0(t) Q_0(t)) \in \mathcal B(\mathcal H)$ be the adiabatic kernel \cite{messiah,nenciu,nenciu2,joye} and $V(t) \in \mathcal U(\mathcal H)$ be the intertwining operator \cite{messiah,nenciu,nenciu2,joye} defined by $\ihbar \dot V(t) = K(t)V(t)$ and satisfying $V(t)P_0(0) = P_0(t)V(t)$. The modified reduced wave operator $Y(t) = V(t)^{-1} X(t) V(t)$ satisfies the following equation
\begin{equation}
\label{eqY}
\ihbar \dot Y(t) = (Q_0(0) - Y(t)) \tilde H_{adiab}(t) (P_0(0) + Y(t))
\end{equation}
with $Q_0(0) Y(t) = Y(t)$, $Y(t)P_0(0) = Y(t)$. $\tilde H_{adiab}(t)$ is the adiabatic renormalization of the Hamiltonian:
\begin{equation}
\tilde H_{adiab}(t) = V(t)^{-1} (H(t) - K(t)) V(t)
\end{equation}
Such a renormalized Hamiltonian occurs in the demonstrations of the strict adiabatic theorems \cite{messiah,nenciu,nenciu2,joye}. The proof of equation (\ref{eqY}) can be found in \ref{appB}. A very interesting fact is that equation (\ref{eqY}) is exactly \cite{jolicard} the equation of an usual time-dependent reduced wave operator $Y(t)$ associated with the Hamiltonian $\tilde H_{adiab}(t)$ and with the fixed active space $\Ran P_0(0)$. To compute $Y(t)$ we can then use a differencing scheme of integration \cite{jolicard,jolicard2} or use the RDWA algorithm in the Floquet extended Hilbert space for the Bloch equation $(\tilde H_{adiab}(t)-\ihbar \partial_t) \Upsilon(t) = \Upsilon(t) (\tilde H_{adiab}(t)-\ihbar \partial_t) \Upsilon(t)$ with $\Upsilon(t) = P_0(0) + Y(t)$. The intertwining operator $V(t)$ is purely geometric and can be calculated by using generalized geometric phases (see \ref{appB4}).

\section{Non-hermitian quantum dynamics surrounding an exceptional point}
Let $\vec R \mapsto H(\vec R)$ be a parameter dependent non-hermitian hamiltonian (with eigenvalues in lower complex halfplane), the set of all possible parameters $\{\vec R\}$ forming a manifold $M$. An exceptional point $\vec R_* \in M$ is a point of coalescence of two non-degenerate complex eigenvalues of $H$ where $H(\vec R_*)$ is not diagonalizable (in contrast with a diabolic point). Let $\mathcal C$ be a closed path in $M$ surrounding $\vec R^*$ (and not other coalescence points). Starting from an eigenvector associated with one of the coalescent eigenvalues, if $\mathcal C$ is slowly followed, the adiabatic approximation states that after one turn the wave function is projected only on the eigenvector associated with the other coalescent eigenvalue \cite{mailybaev,mehri}. This inversion of state is not due to non-adiabatic transitions but it is a topological effect. Using this effect it is possible to propose mechanisms of molecular vibrational cooling or logical gate for quantum information process. But recently, some authors \cite{jaouadi, leclerc} have shown that the conditions to respect the adiabatic approximation are very drastic and need a very slow travelling speed of $\mathcal C$. If we study the mathematically proved adiabatic theorems for non-selfadjoint hamiltonians \cite{nenciu2,joye}, we show that an assumption to be adiabatic is that the wave function remains projected onto the eigenvector associated with the less dissipative eigenvalue. Such an assumption is not satisfied for a path surrounding an exceptional point precisely because of the wanted effect of state topological inversion. This assumption is required because of a competition between adiabatic and dissipative processes. This problem is an interesting area to test the almost adiabatic formalism. This is the subject of this section. We propose two examples of this situation. The first one consists to a non-hermitian two level system which can be viewed as the simplest model exhibiting an exceptional point. This analytical model permits to enlighten the behaviour of the almost adiabatic representation. But in order to show the interest of the almost adiabatic approach in numerical simulations, we need a second example. We consider the problem originally treated in \cite{jaouadi, leclerc} of the coalescence of two instantaneous eigenvalues of the molecular ion $H_2^+$.

\subsection{First example: the two level system}
\subsubsection{The model:}
\begin{figure}
\begin{center}
\includegraphics[width=10cm]{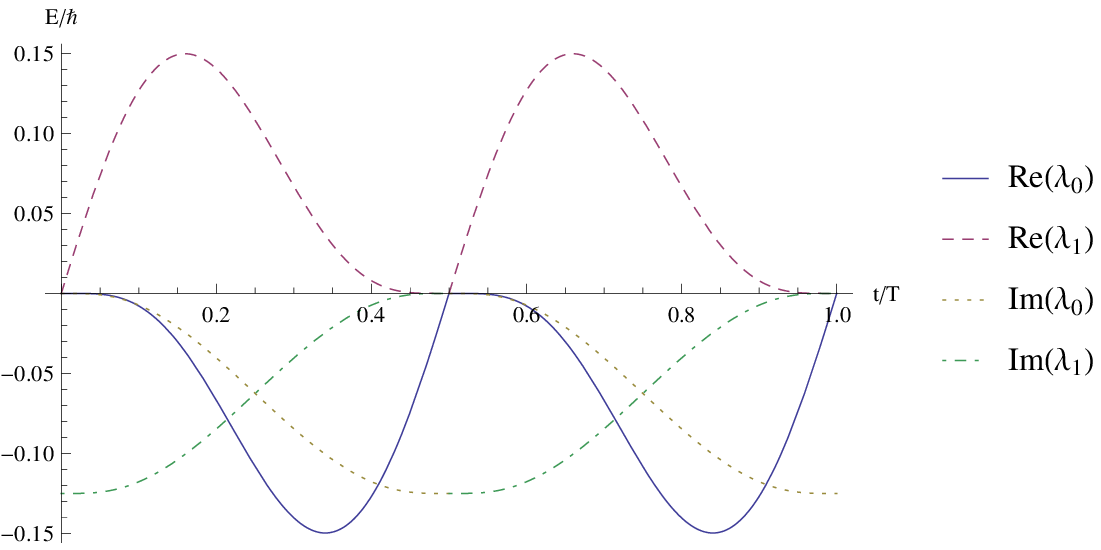}
\caption{\label{eigenvalues} $\lambda_0(t)$ and $\lambda_1(t)$ for the path $\mathcal C$ surrounding two times the exceptional point.}
\end{center}
\end{figure}

We consider the two-level system governed by the Hamiltonian (in a basis denoted by $(|0\rangle,|1\rangle)$)
\begin{equation}
H(\vec R) = \frac{\hbar}{2} \left( \begin{array}{cc} 0 & W \\ W & 2 \Delta - \imath \frac{\Gamma}{2} \end{array} \right) 
\end{equation}
with $\vec R = (W,\Delta) \in M=\mathbb R^+ \times \mathbb R$ and $\Gamma = 0.5$ atomic unit is a constant. This Hamiltonian corresponds to a two-level atom interacting with a laser field in the rotating wave approximation, where $W = |\langle 0|\vec \mu \cdot \vec {\mathcal E} |1 \rangle|$ ($\vec \mu$ is the atomic electric dipole moment and $\vec {\mathcal E}$ is the electric field) and $\Delta = \omega_{01} - \omega_l$ (where $\omega_{01}$ is the Rabi frequency of the transition from $|0\rangle$ to $|1 \rangle$ and $\omega_l$ is the laser frequency). The resonance width $\Gamma$ could modelize a coupling of the state $|1 \rangle$ with the ionization continuum of the atom or a spontaneous emission decay. The interest of considering such a small system, is that it exists analytical expressions of the eigenvectors and of the geometric phase generators. In contrast with a system governed by a large dimensional hamiltonian, no other approximation is needed in addition to adiabatic approximations and/or time propagation schemes. This will permit an unambiguous comparison between different representations of the dynamics. The eigenvalues are
\begin{eqnarray}
\lambda_0 & = & \frac{\hbar}{2} ( z - \sqrt{W^2+z^2})  \\
\lambda_1 & = & \frac{\hbar}{2} ( z + \sqrt{W^2+z^2}) 
\end{eqnarray}
where $z=\Delta - \imath \frac{\Gamma}{4}$. $\vec R_* = (\frac{\Gamma}{4},0)$ is an exceptional point. We set $w = \sqrt{W^2+z^2}$. These eigenvalues are associated with the following eigenvectors:
\begin{equation}
\phi_0^+ = \left( \begin{array}{c}  z + w \\ - W \end{array} \right) \quad,\quad \phi_1^+ = \left( \begin{array}{c} W \\ z + w  \end{array}  \right)
\end{equation}
for the Riemann sheet such that $\sqrt{z^2} = z$ and
\begin{equation}
\phi_0^- = \left( \begin{array}{c} W \\ z - w \end{array} \right) \quad,\quad \phi_1^- = \left( \begin{array}{c}  z - w  \\- W \end{array}  \right)
\end{equation}
for the Riemann sheet such that $\sqrt{z^2}=-z$. The following of a closed path surrounding $\vec R_*$ induces a passage from a Riemann sheet to another one. We can note that this question is related to the labelling procedure of the states \cite{viennot5}. With the present labelling, the imaginary parts of the eigenvalues are not continuous by the passage through $[0, \frac{\Gamma}{4}] \times \{0\}$. Let $\mathcal C$ be the path defined as being a circle centered on $\vec R_*$, starting from $(0,0)$ (laser field is off) and surrounding two times $\vec R_*$:
\begin{eqnarray}
W(t) & = & \frac{\Gamma}{4} (1-\cos(4\pi \frac{t}{T}))  \\
\Delta(t) & = & \frac{\Gamma}{4} \sin (4\pi \frac{t}{T}) 
\end{eqnarray}
where $T$ is the duration of the interaction. Figure \ref{eigenvalues} shows the evolution of the eigenvalues when this path is followed.
$H$ being not self-adjoint, the eigenvector set is not orthonormal but biorthonormal to $(\langle \phi^{\pm}_0*|,\langle \phi^{\pm}_1*|)$ with
\begin{equation}
\langle \phi^{+}_0*| = \frac{1}{2w(w+z)} (z+w \quad -W)
\end{equation}
\begin{equation}
\langle \phi^{-}_0*| = \frac{1}{2w(w-z)} (W \quad z-w )
\end{equation}
\begin{equation}
\langle \phi^+_1* | = \frac{1}{2w(w+z)}(W \quad z+w)
\end{equation}
\begin{equation}
\langle \phi^-_1* | = \frac{1}{2w(w-z)}(z-w \quad -W )
\end{equation}
with
\begin{equation}
\langle \phi_i^\pm*|\phi_j^\pm \rangle = \delta_{ij}
\end{equation}
The generator of the geometric phase is then
\begin{equation}
A^\pm = \langle \phi^\pm_0*|\partial_t|\phi^\pm_0 \rangle = \langle \phi^\pm_1*|\partial_t|\phi^\pm_1 \rangle = \frac{(2w\pm z)W \dot W\pm(w\pm z)^2 \dot z}{2w^2(w\pm z)}
\end{equation}
and the non-adiabatic coupling term is
\begin{equation}
\mathcal A^\pm_{10} = \langle \phi^\pm_1*|\partial_t|\phi^\pm_0\rangle = \pm \frac{z \dot W - W \dot z}{2w^2}
\end{equation}
(with also $\mathcal A^\pm_{01} = - \mathcal A^\pm_{10}$).

\subsubsection{The dynamics:}
\begin{figure}
\begin{center}
\includegraphics[width=10cm]{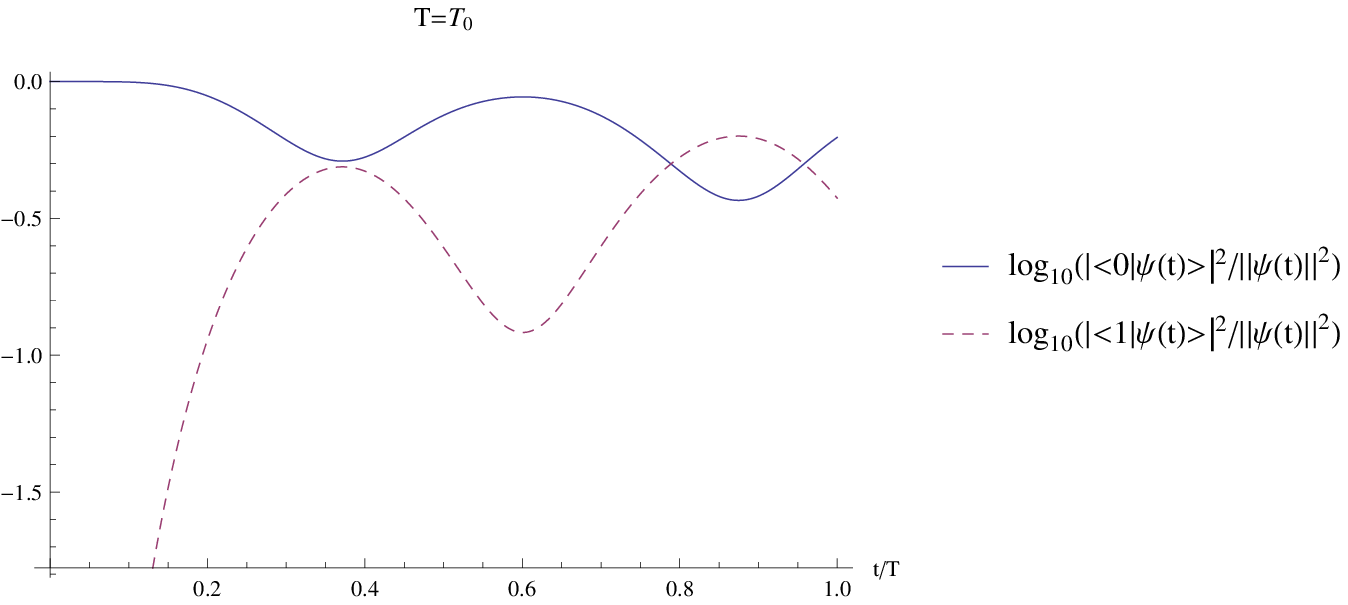}
\includegraphics[width=10cm]{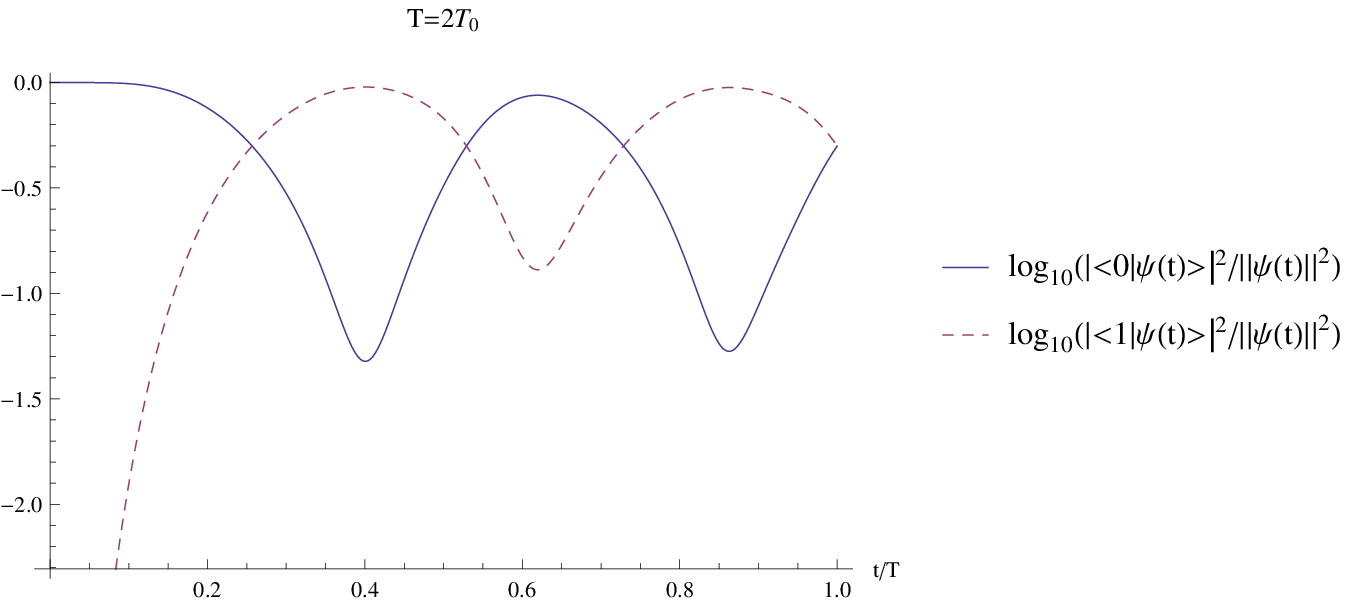}\\
\includegraphics[width=10cm]{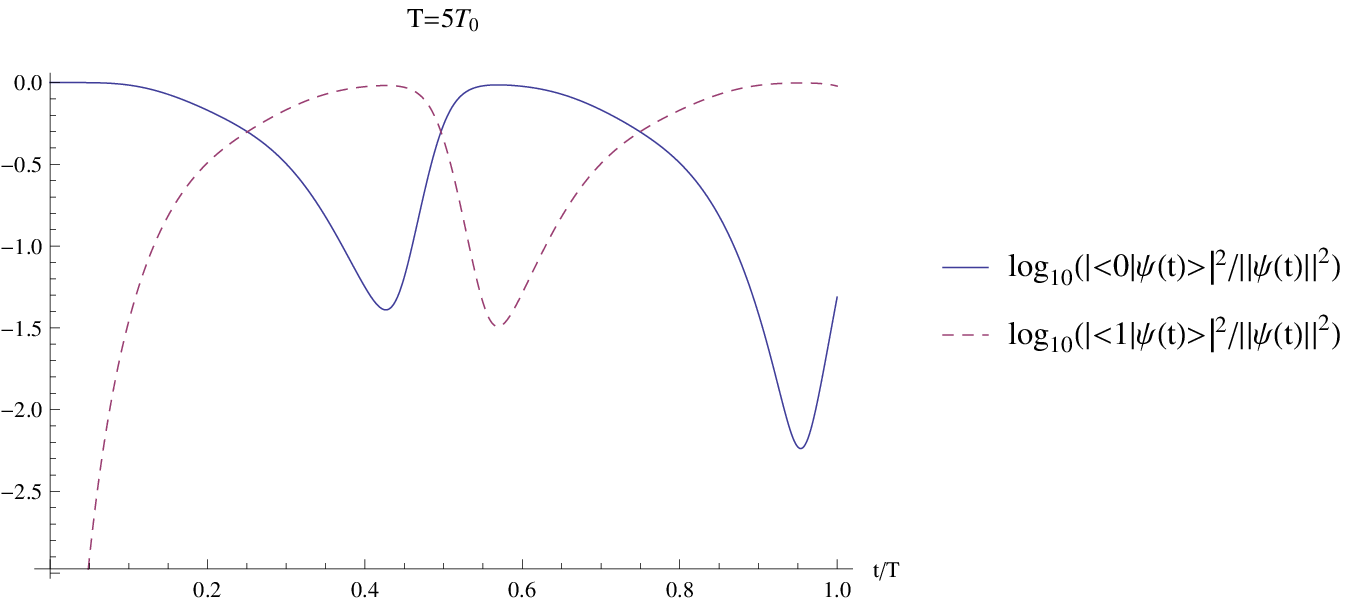}
\includegraphics[width=10cm]{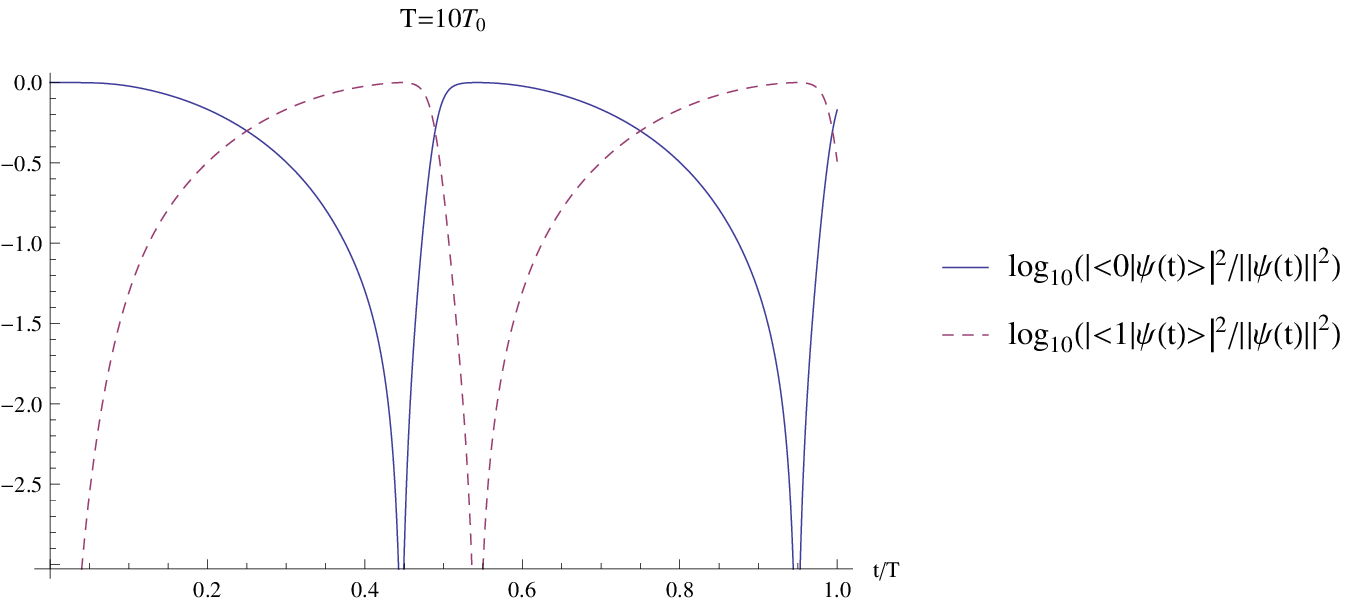}
\caption{\label{populations} Evolution of the populations for different evolution speeds.}
\end{center}
\end{figure}

We start with $\psi(0) = \phi_0^+(0) = |0 \rangle = {1 \choose 0}$ (the less dissipative state), and we consider the time-dependent wave function $\psi(t)$ solution of the Schr\"odinger equation $\ihbar \dot \psi = H(W(t),\Delta(t)) \psi(t)$. We want to compare a strict adiabatic approximation and an almost adiabatic approximation with the true wave function. We need then a ``numerically exact'' solution of the Schr\"odinger equation. To that, we use a splitted second order differencing propagation scheme (see \ref{appC}). It is a mixing of a split operator method for the non-hermitian part (ensuring a correct description of the dissipation process) and of a second order differencing scheme (ensuring a sufficient accuracy of the propagation) \cite{leforestier}.\\
Let $T_0= 50$ atomic unit of time. We consider the dynamics for a very fast following of the path $\mathcal C$ with $T=T_0$, a very slow following of the path $\mathcal C$ with $T=10 T_0$, and intermediate cases with $T=2 T_0$ and $T=5 T_0$. The renormalized populations $ \frac{|\langle 0|\psi(t)\rangle|^2}{\|\psi(t)\|^2}$ and $ \frac{|\langle 1|\psi(t)\rangle|^2}{\|\psi(t)\|^2}$ are shown in figure \ref{populations}.
We renormalize the populations in order to compare the different cases independently on the falls induced by the dissipation. For $T=10 T_0$ the dynamics is adiabatic and as predict by the adiabatic analysis, after one turn ($t=\frac{T}{2}$) we have a state inversion (and another one during the second turns). We have a less perfect inversion for $T=5 T_0$. But for these two cases, since the dynamics is very slow, the dissipation of the wave function is very strong as we can see in table \ref{dissipation}.
\begin{table}
\begin{center}
\caption{\label{dissipation} Dissipation rates at the end of the evolution.}
\begin{tabular}{c|c}
$T$ & $\log_{10} \|\psi(T)\|^2$ \\
\hline
$T_0$ & $-1.6044$ \\
$2 T_0$ & $-5.5846$ \\
$5 T_0$ & $-13.4441$ \\
$10 T_0$ & $-26.1001$ \\
\end{tabular}
\end{center}
\end{table}
The adiabatic inversions are realized but the system is completely killed by the dissipation. For acceptable dissipations ($T=T_0$ or $T=2 T_0$), the adiabatic inversions are not correctly realized.

\subsubsection{Representations of the dynamics:}
\begin{figure}
\begin{center}
\includegraphics[width=10cm]{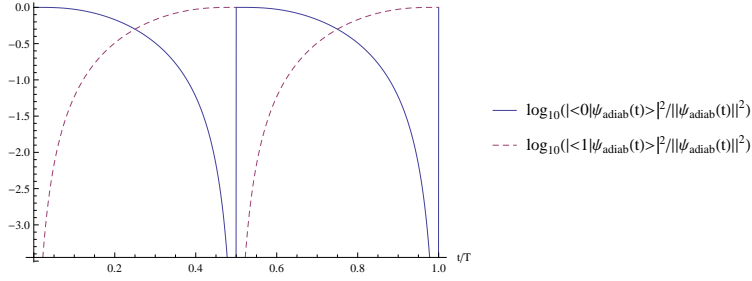}
\caption{\label{popadiab} Evolution of the populations in the adiabatic approximation (this graph is independent from the speed of the evolution).}
\end{center}
\end{figure}

\begin{figure}
\begin{center}
\includegraphics[width=10cm]{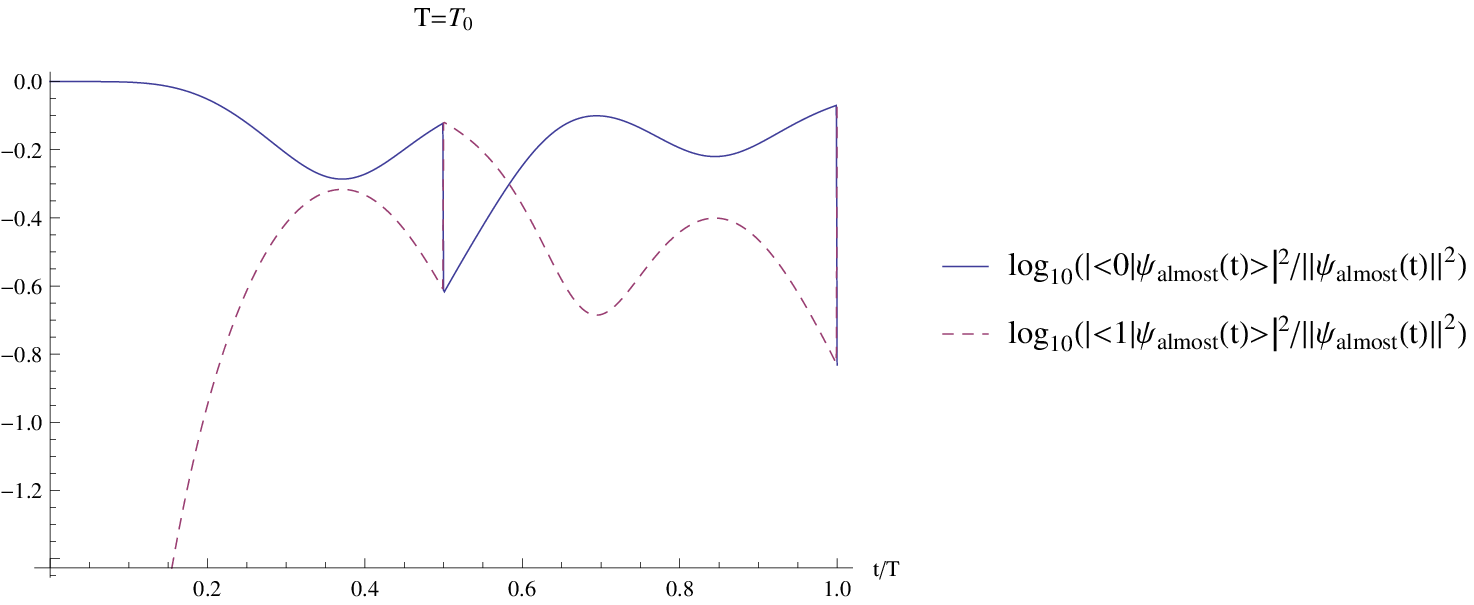}
\includegraphics[width=10cm]{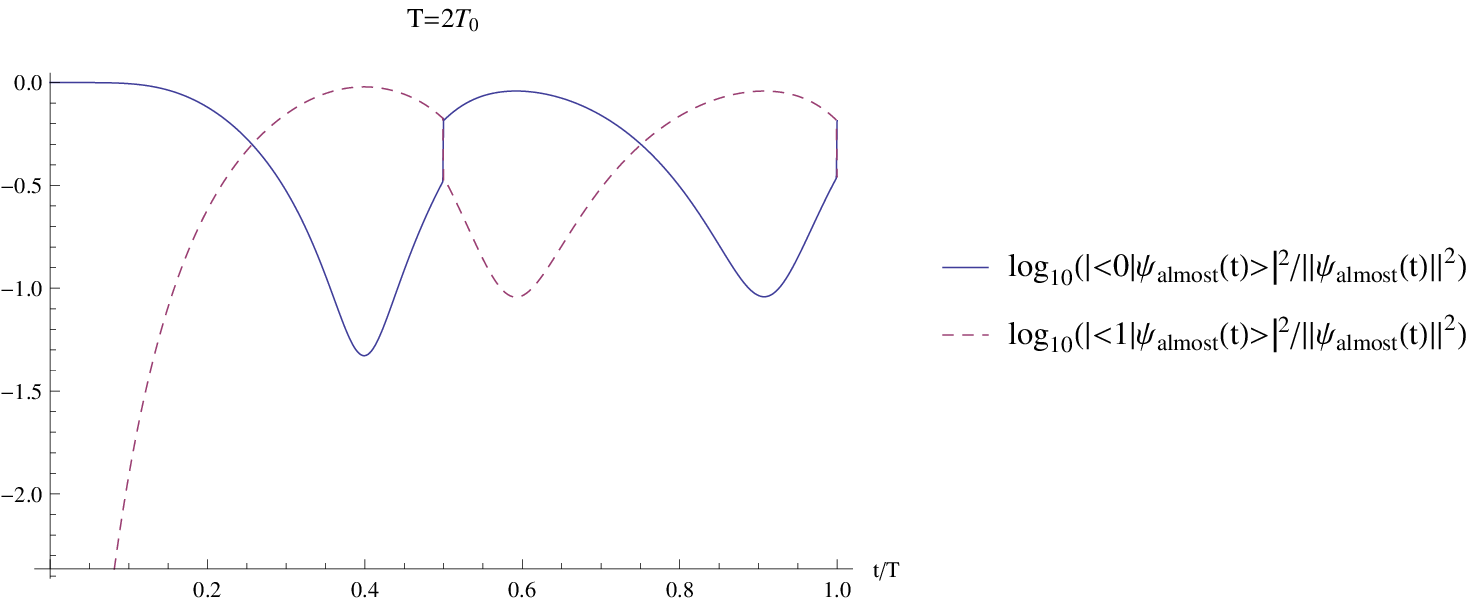}\\
\includegraphics[width=10cm]{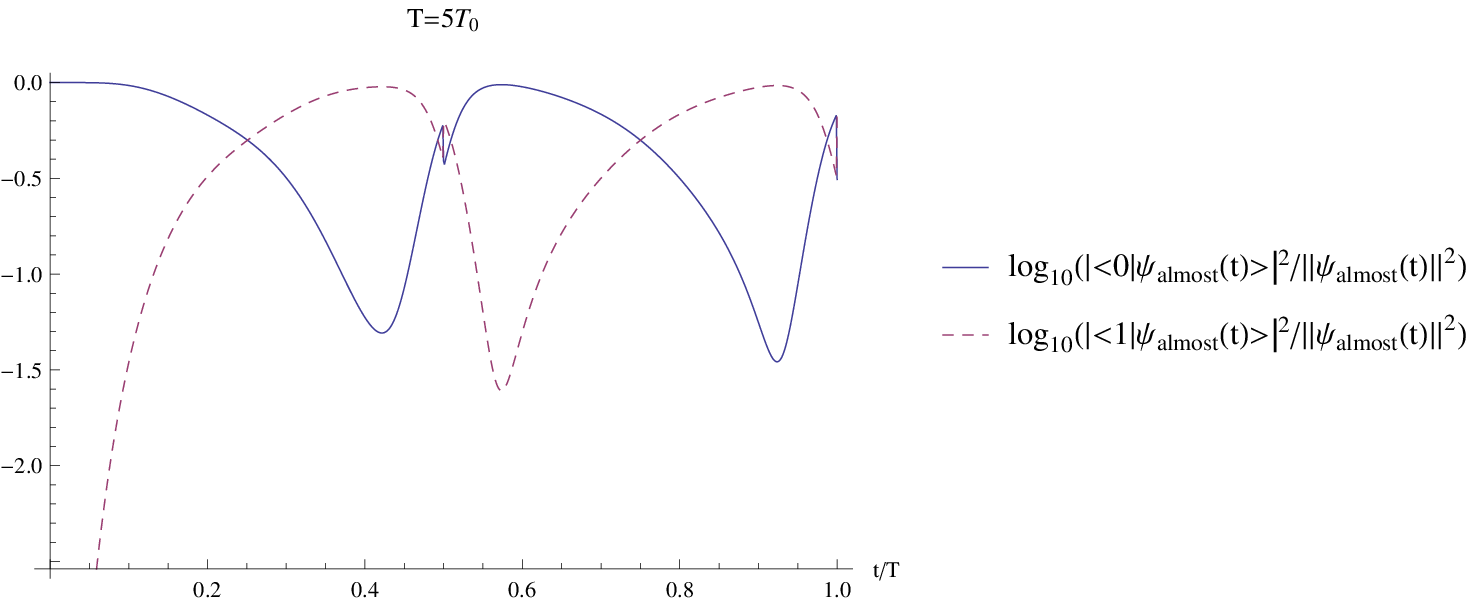}
\includegraphics[width=10cm]{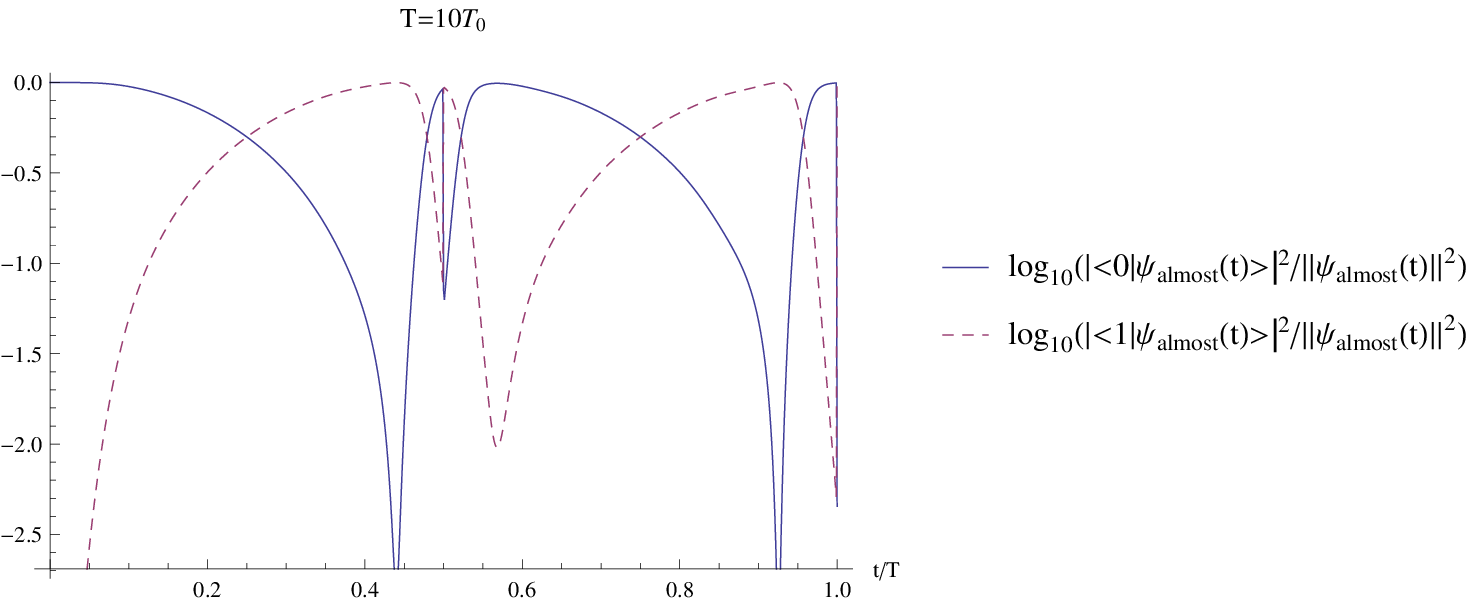}
\caption{\label{popWO} Evolution of the populations for different evolution speeds computed in the almost adiabatic representation with a first order differencing scheme to compute the wave operator.}
\end{center}
\end{figure}

\begin{figure}
\begin{center}
\includegraphics[width=10cm]{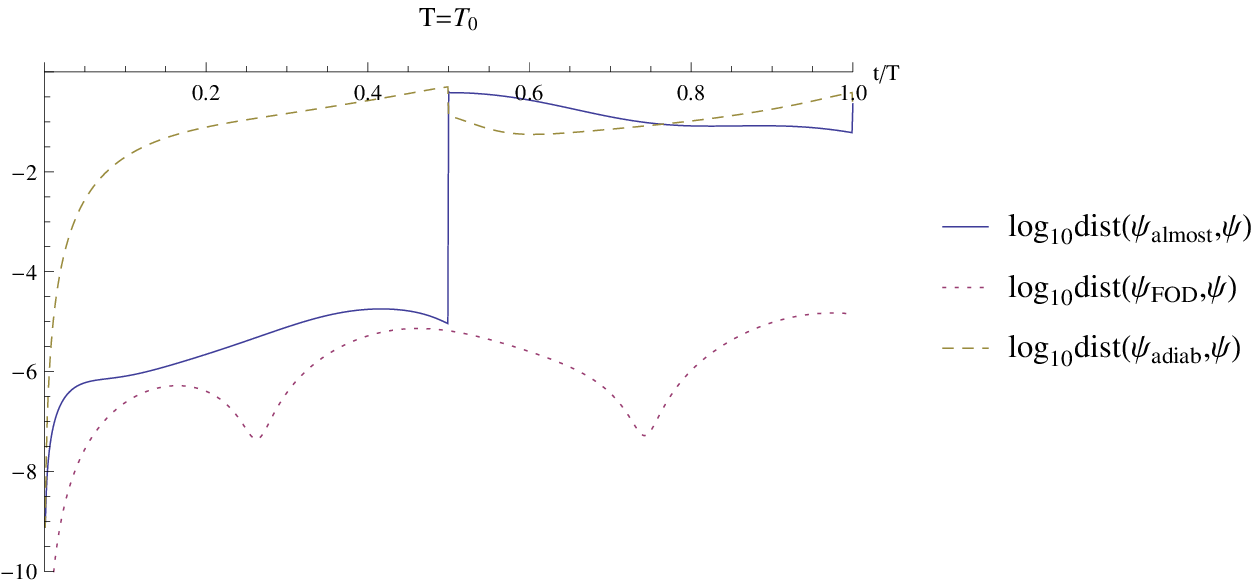}
\includegraphics[width=10cm]{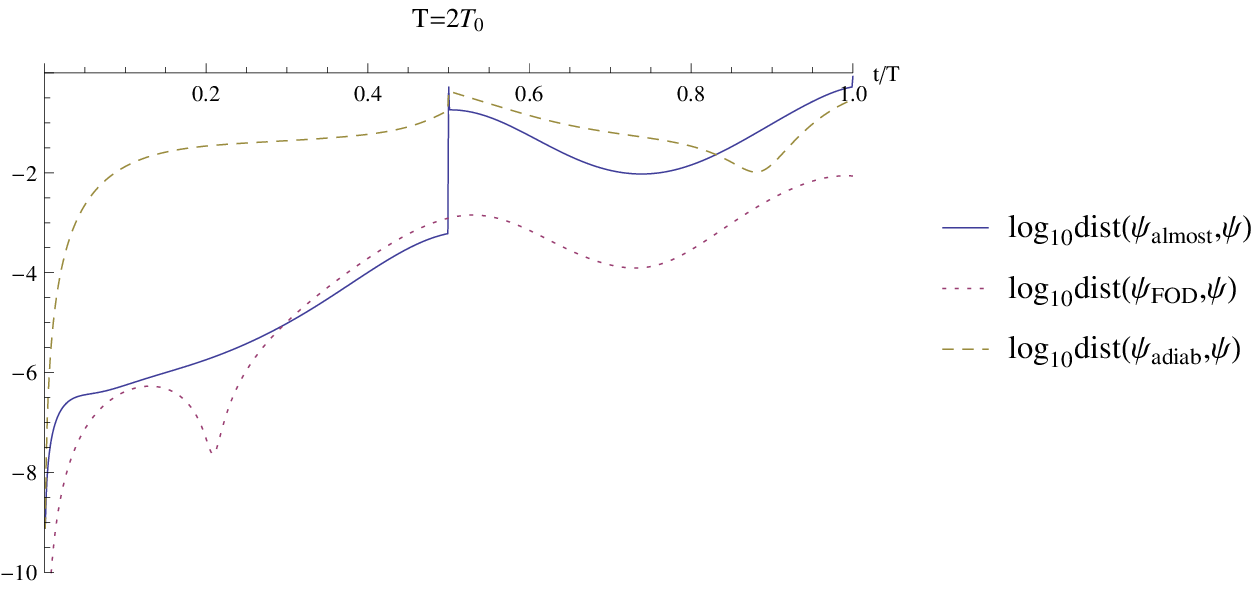}\\
\includegraphics[width=10cm]{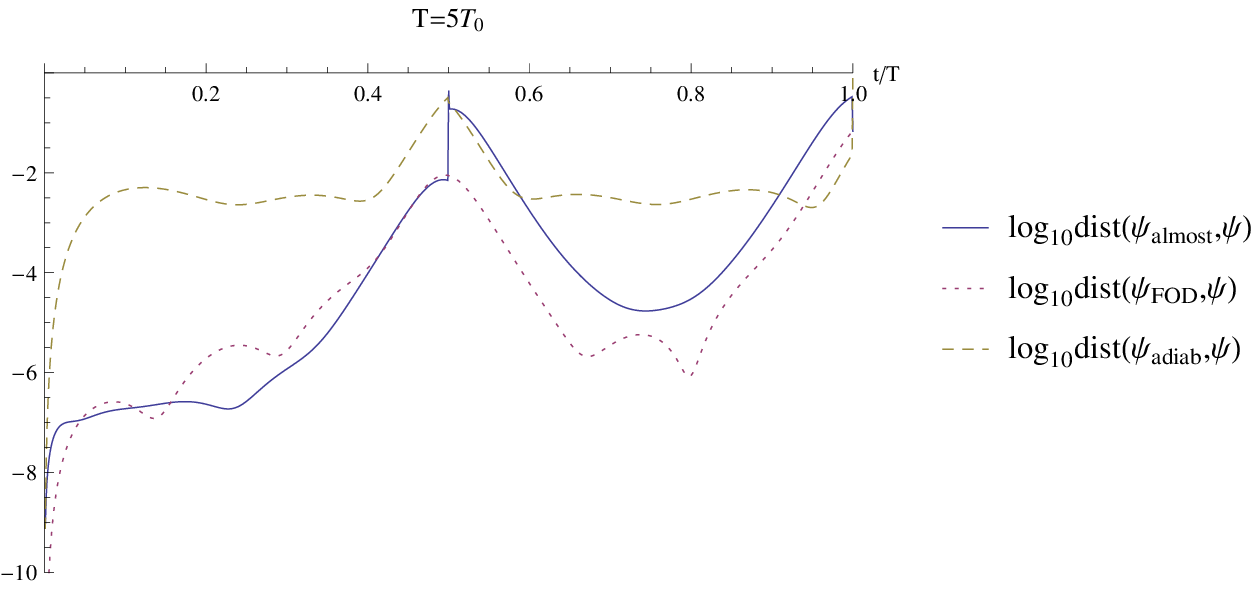}
\includegraphics[width=10cm]{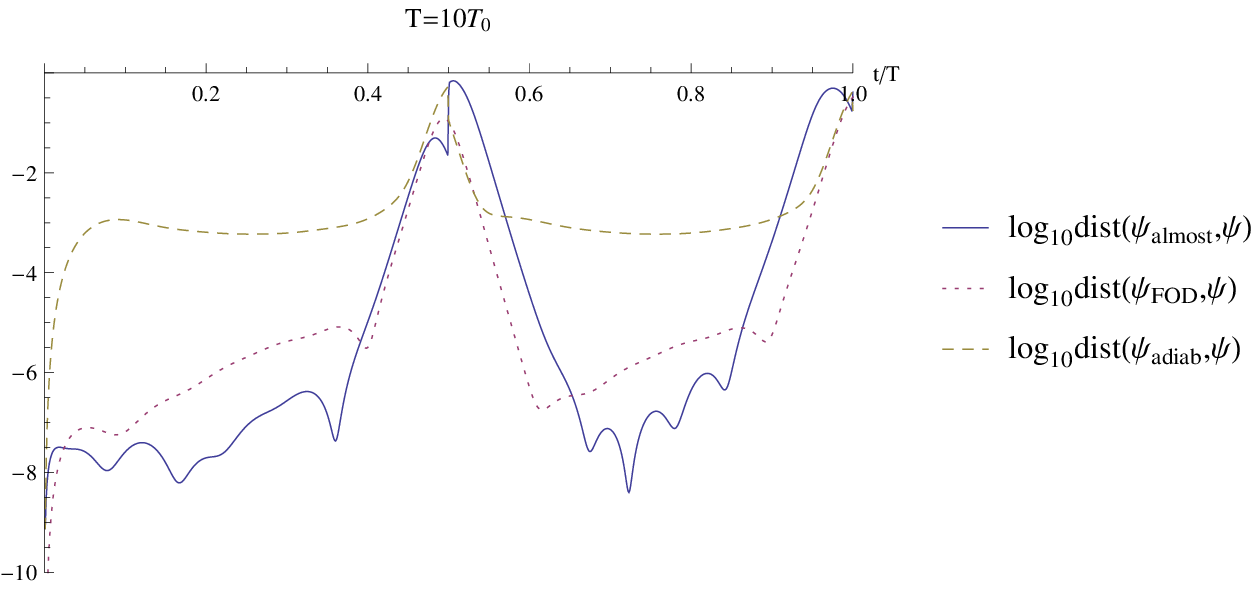}
\caption{\label{dist} Evolution of the error concerning the evaluation of the wave function for the almost adiabatic representation, the adiabatic approximation, and a non-splitting first order differencing scheme of integration of the Schr\"odinger equation. The distance is defined by $\mathrm{dist}(\psi,\phi) = 1 - \frac{|\langle \psi|\phi \rangle|}{\|\psi\| \|\phi\|}$.}
\end{center}
\end{figure}

\begin{figure}
\begin{center}
\includegraphics[width=10cm]{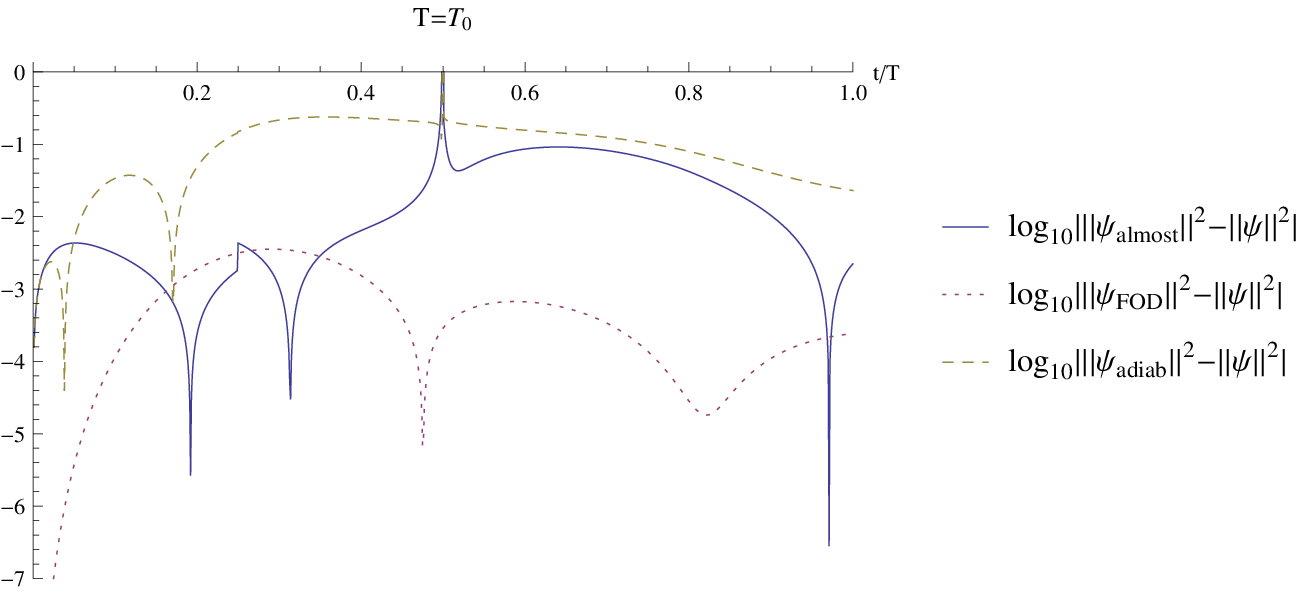}
\includegraphics[width=10cm]{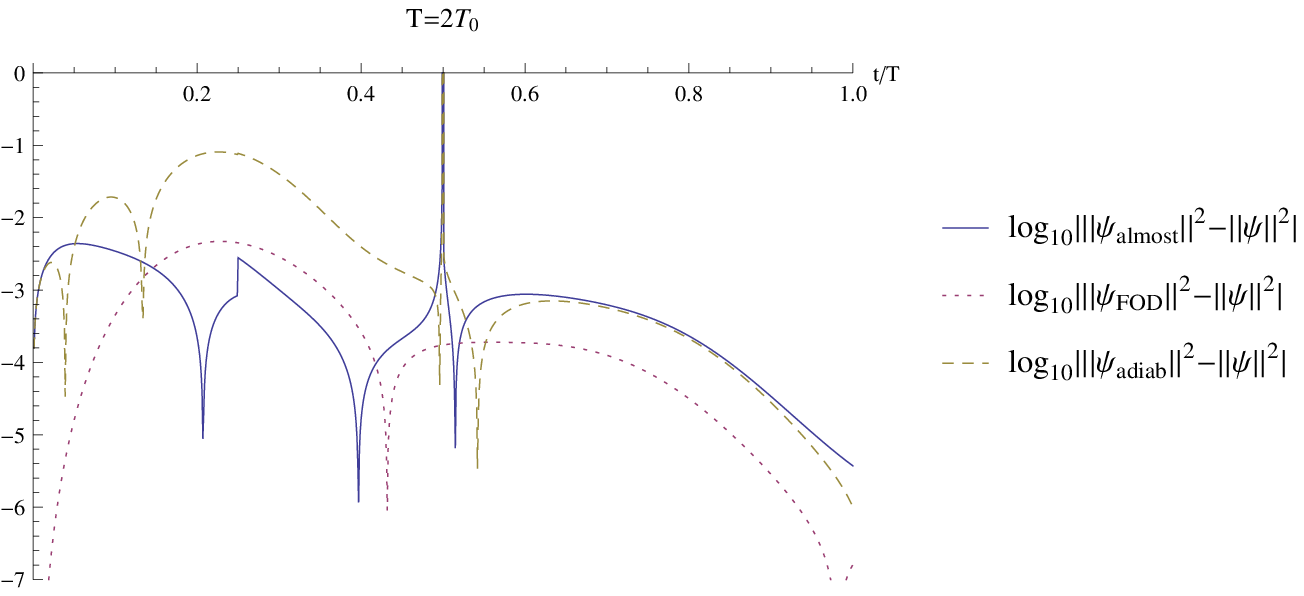}\\
\includegraphics[width=10cm]{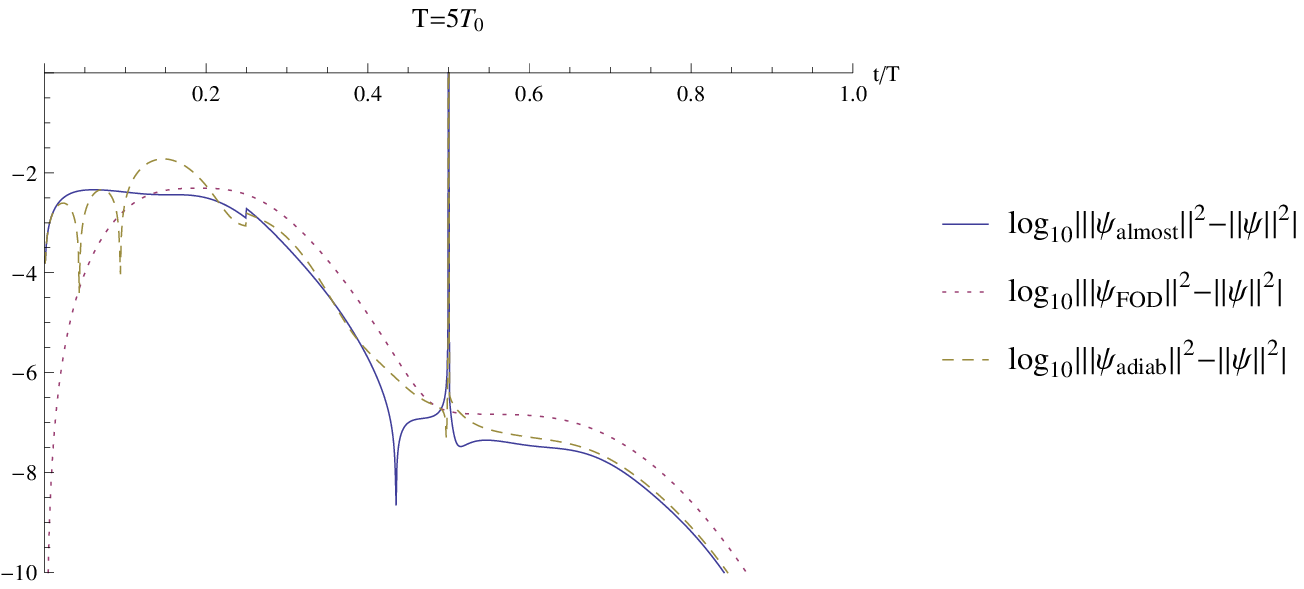}
\includegraphics[width=10cm]{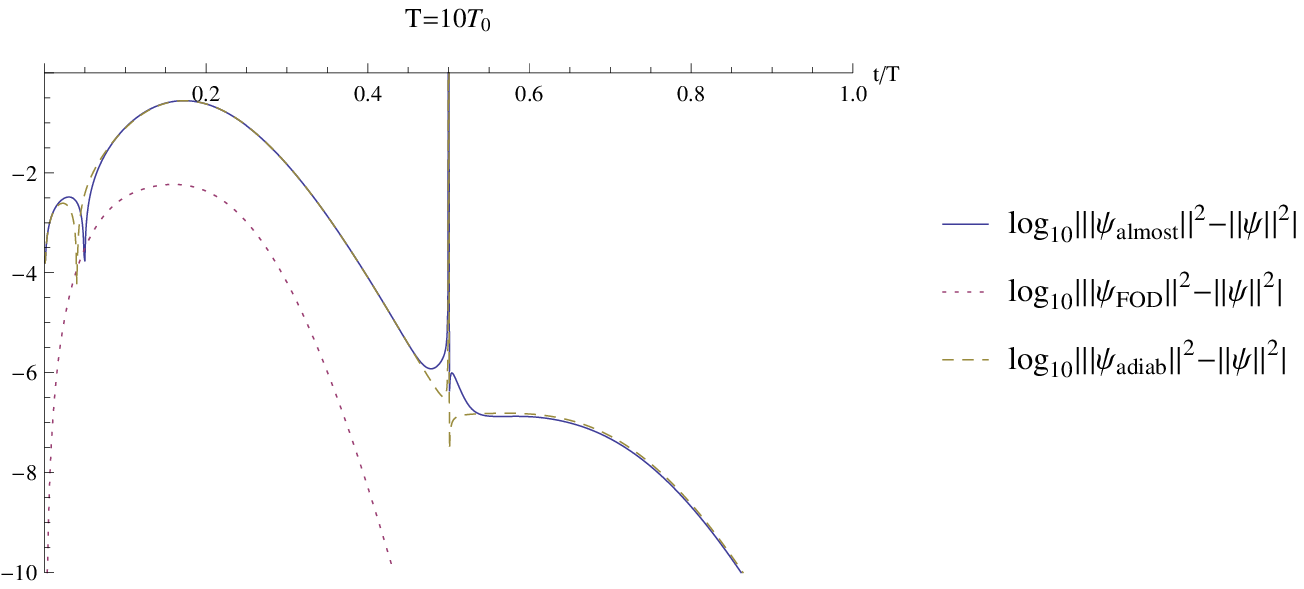}
\caption{\label{errnorm} Evolution of the error concerning the normalization for the almost adiabatic representation, the adiabatic approximation, and a non-splitting first order differencing scheme of integration of the Schr\"odinger equation.}
\end{center}
\end{figure}

\begin{figure}
\begin{center}
\includegraphics[width=10cm]{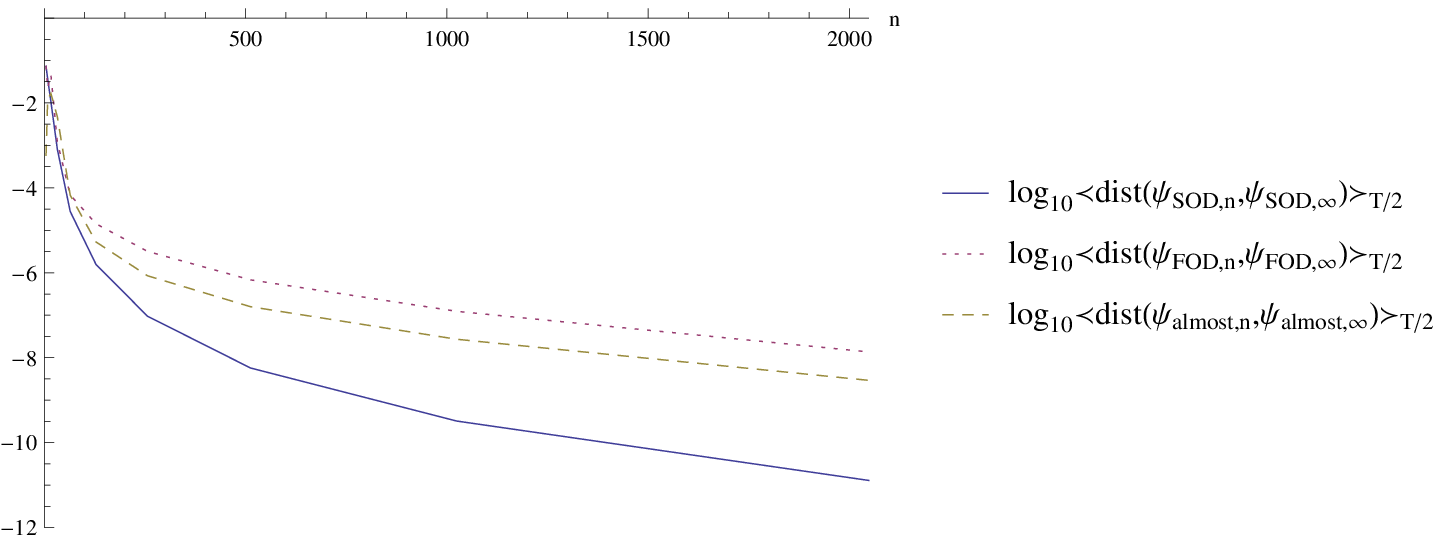}
\caption{\label{convergatom} Convergences of the second order differencing scheme (SOD), the first order differencing scheme (FOD) and the almost adiabatic representation in a logarithmic scale with respect to $n$ the number of time steps using in the propagation schemes during $[0,T]$. The convergence accuracies are computed with respect to reference solutions denoted $\psi_{*,\infty}$ computed with $4000$ time steps (the graphics presented here do not change significantly with respect to reference solutions computed with sufficiently large number of time steps). The convergence accuracy for $n$ time steps is computed as being $ \prec \mathrm{dist}(\psi_{*,n},\psi_{*,\infty}) \succ_{T/2} = \frac{2}{T} \int_0^{T/2} \mathrm{dist}(\psi_{*,n}(t),\psi_{*,\infty}(t)) dt$ with $\mathrm{dist}(\psi,\phi) = 1 - \frac{|\langle \psi|\phi\rangle|}{\|\psi\| \|\phi\|}$. The figures previously presented are computed with $1000$ time steps, corresponding to a ``numerically exact'' solution converged with an accuracy close to $10^{-10}$ and an almost adiabatic representation converged with an accuracy close to $10^{-7}$.}
\end{center}
\end{figure}

The formulea of the strict and almost adiabatic representations can be generalized without difficulty by taking into account the biorthonormality (we have $P_0 = |\phi_0^+(t)\rangle \langle \phi_0^+(t)*|$ in place of the orthogonal projection, and the Schr\"odinger-von Neumann equation \ref{SvN} becomes $\ihbar \dot P(t) = [H(t),P(t)]_+ = H(t)P(t)+P(t)H(t)$ where $P(t) = |\psi(t)\rangle \langle \psi^*(t)|$ is the projection on the solution of the Schr\"odinger equation $\ihbar \dot \psi(t) = H(t) \psi(t)$ parallel to the orthogonal supplement of the space spaned by the solution of the equation $-\ihbar \dot \psi^*(t) = H(t)^\dagger \psi^*(t)$ -- the star does not denote the complex conjugation but the biorthogonality with $\psi$ --). In the strict adiabatic approximation, the wave function is represented by
\begin{eqnarray}
& & \psi_{adiab}(t) \nonumber \\
& & = \left\{ \begin{array}{cc} e^{-\ihbar^{-1}\int_0^t \lambda_0(t')dt' - \int_0^t A^+(t')dt'} \phi_0^+(t)  & \text{if } t \leq \frac{T}{2} \\
e^{-\ihbar^{-1}\int_0^t \lambda_0(t')dt' - \int_0^{T/2} A^+(t')dt' - \int_{T/2}^t A^-(t')dt'} \phi_0^-(t)  & \text{if } t > \frac{T}{2} \end{array} \right.
\end{eqnarray}
where we have taken into account the change of Riemann sheet after one turn. The populations evaluated with this formula are shown in figure \ref{popadiab}.
As expected, the adiabatic approximation is not correct for $T=T_0$ and $T=2T_0$. We consider the almost adiabatic representation:
\begin{eqnarray}
& & \psi_{almost}(t) = \nonumber \\
& & e^{-\ihbar^{-1}\int_0^t \lambda_0(t')dt' - \int_0^t A^+(t')dt'- \int_0^t \eta(t')dt' } (\phi_0^+(t) + x(t) \phi_1^+(t))  \text{ if } t \leq \frac{T}{2}; \nonumber \\
& & e^{-\ihbar^{-1}\int_0^t \lambda_0(t')dt' - \int_0^{T/2} A^+(t')dt' - \int_{T/2}^t A^-(t')dt'- \int_0^t \eta(t')dt' } \nonumber \\
& & \qquad \qquad \times (\phi_0^-(t) + x(t) \phi_1^-(t))  \text{ if } t > \frac{T}{2}
\end{eqnarray}
where we have written the wave operator as being
\begin{equation}
\Omega(t) = \left\{\begin{array}{cc} |\phi_0^+(t) \rangle \langle \phi_0^+(t)*| + x(t) |\phi_1^+(t) \rangle \langle \phi_0^+(t)*|  & \text{if } t \leq \frac{T}{2} \\
|\phi_0^-(t) \rangle \langle \phi_0^-(t)*| + x(t) |\phi_1^-(t) \rangle \langle \phi_0^-(t)*| &  \text{if } t > \frac{T}{2} \end{array} \right.
\end{equation}
and
\begin{equation}
\eta(t) = \left\{\begin{array}{cc} \langle \phi_0^+(t)*|\dot \Omega(t) | \phi_0^+(t) \rangle  & \text{if } t \leq \frac{T}{2} \\
 \langle \phi_0^-(t)*|\dot \Omega(t) | \phi_0^-(t) \rangle  & \text{if } t > \frac{T}{2} \end{array} \right.
\end{equation}
The renormalized hamiltonian is
\begin{equation}
\tilde H_{adiab} = \left(\begin{array}{cc} \lambda_0 & - \ihbar \mathcal A^\pm_{10} \\ \ihbar \mathcal A^\pm_{10} & \lambda_1 \end{array} \right)
\end{equation}
and in this example equation \ref{eqY} is reduced to
\begin{equation}
\label{eqx}
\dot x(t) = \left\{\begin{array}{cc} \mathcal A_{10}^+(t) x(t)^2 - \ihbar^{-1} (\lambda_1(t)-\lambda_0(t))x(t) + \mathcal A^+_{10}(t)  & \text{if } t\leq \frac{T}{2} \\
\mathcal A_{10}^-(t) x(t)^2 - \ihbar^{-1} (\lambda_1(t)-\lambda_0(t))x(t) + \mathcal A^-_{10}(t)  & \text{if } t> \frac{T}{2} \end{array} \right.
\end{equation}
and we have $\eta(t) = x(t) \mathcal A_{10}^\pm(t)$. In the almost adiabatic representation, the approximation is present in the integration of equation \ref{eqx}. We chose a first order differencing scheme for the propagation:
\begin{equation}
x_{n+1} = x_n + \left(\mathcal A_{10}^\pm(t_n) x_n^2 - \ihbar^{-1}(\lambda_1(t_n)-\lambda_0(t_n))x_n + \mathcal A_{10}^\pm(t_n) \right) \Delta t
\end{equation}
with samer step $\Delta t$ as for the ``numerically exact'' solution of the Schr\"odinger equation. The populations computed with the almost adiabatic representation are shown figure \ref{popWO}.
A comparison of the figures \ref{populations}, \ref{popadiab} and \ref{popWO} shows:
\begin{itemize}
\item For $T=2T_0$ (non-adiabatic regime), the almost adiabatic representation corrects completely the errors of the adiabatic approximation (except in the neigbourhood of $T/2$ -- the passage from a Riemann sheet to another one -- where the almost adiabatic dynamics is too brutal because of the needed correction to the strict adiabaticity is very strong).
\item For $T=5T_0$ (adiabatic regime), the almost adiabatic representation is in a very good accordance with reality, the deviations to the strict adiabatic approximation are corrected.
\item For $T=T_0$ (very non-adiabatic regime), the almost adiabatic representation corrects the errors of the adiabatic approximation until $T/2$, after the Riemann sheet change, the error of the adiabatic approximation being very strong, the almost adiabatic representation needs some time to correct the populations.
\item For $T=10T_0$ (very adiabatic regime), the almost adiabatic representation seems lower than the adiabatic approximation after the Riemann sheet change. But we note that this corresponds to a very dissipated wave function, and small errors are amplified by the renormalization by $1/\|\psi_{almost}\|^2$.
\end{itemize}
We confirm this analysis by studying the errors. In order to evaluate the effect of the approximation in the calculation of the wave operator, we compare also with a direct integration of the Schr\"odinger equation with a non-splitting first order differencing scheme. The errors concerning the representations of the wave function are drawn figure \ref{dist} and the errors concerning the dissipation are drawn figure \ref{errnorm}.
In all cases, the almost adiabatic representation is clearly better than the adiabatic approximation. At the end of the evolution, the error of the almost adiabatic representation increases. This is certainly caused by the error accumulation associated with the first order differencing scheme used to compute $x(t)$. Concerning the dissipation, the almost adiabatic representation is equivalent to the adiabatic approximation except in non-adiabatic regimes where it is better (only for the first part of the dynamics in the case $T=2T_0$).\\
The convergences of the ``numerically exact'' integration, of the first order differencing scheme and of the almost adiabatic representation are studied figure \ref{convergatom}.
For a very simple example as a two level system, all numerical methods converge and provide results with short numerical computation times. In these conditions, a first order differencing scheme can provide good solutions (as for the example for the very non-adiabatic regime $T=T_0$). But for problems with a large Hilbert space dimension, such a simple numerical scheme does not work. The almost adiabatic representation can provide satisfactory results with relatively short numerical computation times. This is the subject of the next paragraph considering the example of the $H_2^+$ molecule.

\subsection{Second example: $H_2^+$}
\subsubsection{The system:}
\begin{figure}
\begin{center}
\includegraphics[width=10cm]{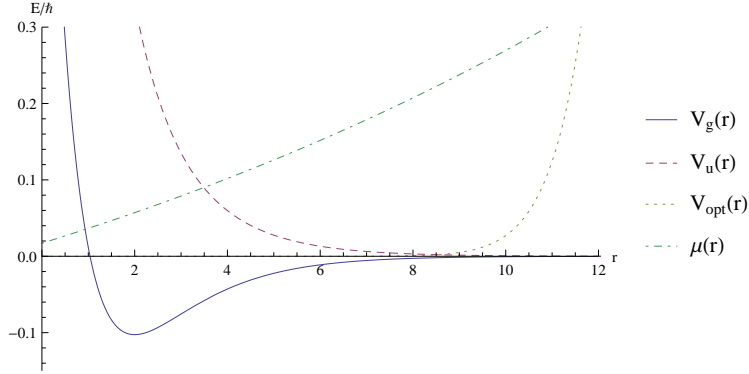}
\caption{\label{potentials} The vibrational potentials for the ground electronic state $V_g(r)$ and for the first excited electronic state $V_u(r)$, the optical potential $V_{opt}(r)$ defining the absorbing boundary and the electric dipole moment $\mu(r)$.}
\end{center}
\end{figure}

\begin{figure}
\begin{center}
\includegraphics[width=7cm]{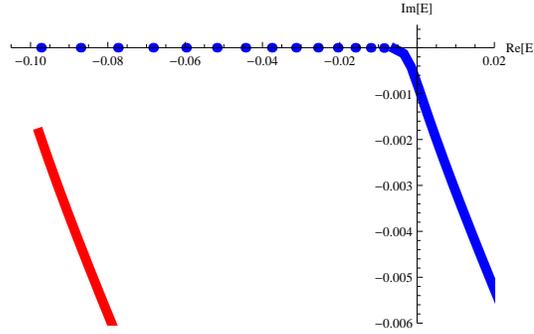}
\caption{\label{spectrum} Spectrum of the Hamiltonian eq. \ref{ham_H2p} for $W=0$ and $\omega= 0.10242$ atomic unit. The spectrum associated with the ground electronic state is represented in blue and the spectrum associated with first excited electronic state is represented in red. The Hamiltonian presents pure point spectrum on the real axis associated with the potential well of the ground vibrational potential and two continuous spectra rotated in the lower complex half plane which are associated with the scattering states of the two vibrational potentials.}
\end{center}
\end{figure}

\begin{figure}
\begin{center}
\includegraphics[width=7cm]{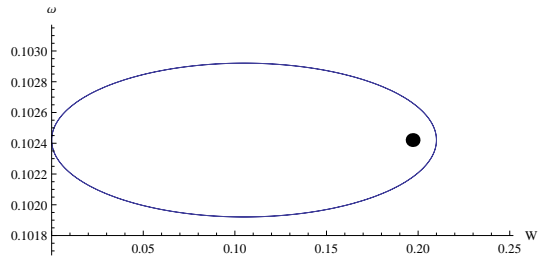}
\caption{\label{path} The path $\mathcal C$ in the manifold $M$. $\bullet$ denotes the exceptional point.}
\end{center}
\end{figure}

\begin{figure}
\begin{center}
\includegraphics[width=10cm]{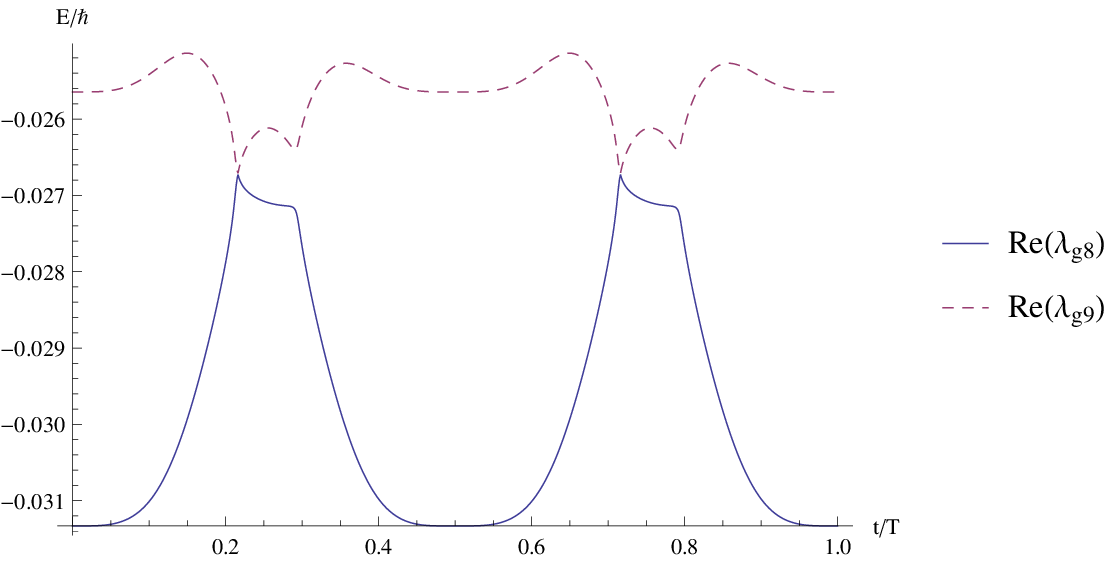} \includegraphics[width=10cm]{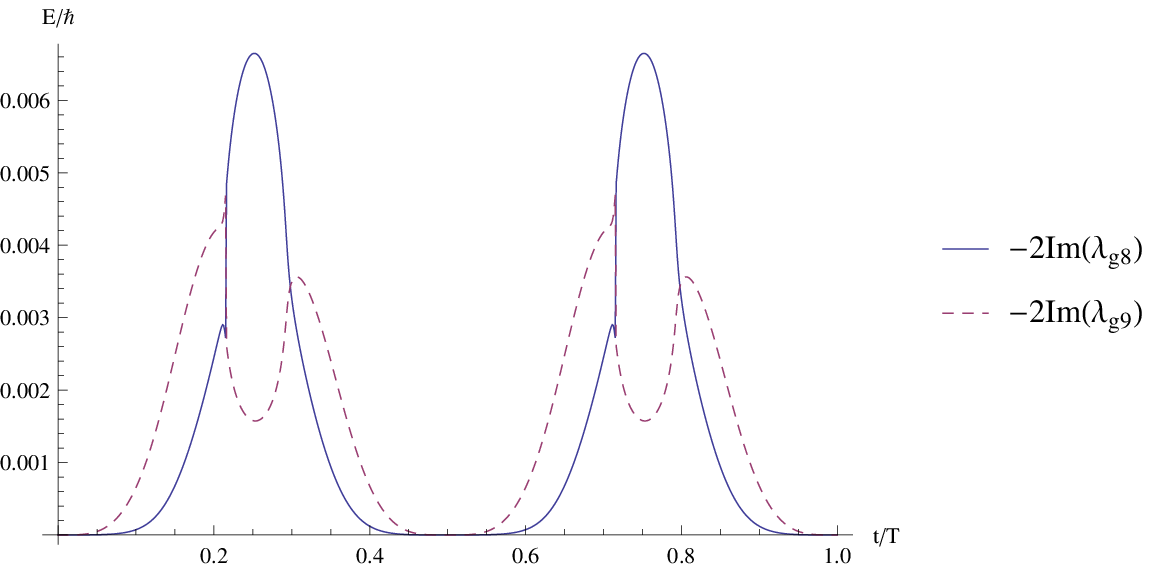}
\caption{\label{eigenvalH2p} $\lambda_{g8}(t)$ and $\lambda_{g9}(t)$ for the path $\mathcal C$ surrounding two times the exceptional point.}
\end{center}
\end{figure}

We consider the vibration of the molecule $H_2^+$ described by the Hilbert space $\mathcal H = L^2(\mathbb R^+,dr) \otimes \mathbb C^2$, where $r$ is the internuclear distance and $\mathbb C^2$ describes the space of the electronic states (we consider only the ground state ${^2}\Sigma^+_g$ and the first excited state ${^2}\Sigma^+_u$). The dynamics of the molecule interacting with a laser field is governed in the rotating wave approximation with one photon by the Hamiltonian
\begin{eqnarray}
\label{ham_H2p}
H(\vec R) & = & - \frac{\hbar^2}{2m} \frac{d^2}{dr^2} \otimes 1_{\mathbb C^2} \nonumber \\
& & \quad + V_g(r) \otimes |{^2}\Sigma^+_g \rangle \langle {^2}\Sigma^+_g| + (V_u(r) - \hbar \omega) |{^2}\Sigma^+_u \rangle \langle {^2}\Sigma^+_u| \nonumber \\
& & \quad + W \mu(r) \otimes (|{^2}\Sigma^+_g \rangle \langle {^2}\Sigma^+_u| + |{^2}\Sigma^+_u \rangle \langle {^2}\Sigma^+_g|) \nonumber \\
& & \quad - \imath V_{opt}(r) \otimes 1_{\mathbb C^2}
\end{eqnarray}
with $\vec R = (W,\omega) \in M = \mathbb R^+ \times \mathbb R^+$. $m=911.389$ atomic unit is the reduced mass of the molecule; $\omega$ is the laser frequency; and $W$ is the electric field. $V_g(r)$ and $V_u(r)$ are the vibrational potentials of the molecule with respect to the electronic state. $\mu(r)$ is the molecular electric dipole moment. The optical potential $-\imath V_{opt}(r)$ plays the role of an absorbing boundary used to dissipate the wave packets going in regions with large $r$. Since numerically it is impossible to describe the infinite configuration space $[0,+\infty[$ for $r$; we must consider only a configuration space $[0,r_{max}]$ (with $r_{max} = 12$ atomic unit). The absorbing boundary avoids unphysical reflexions of the wave packets on the box boundary $r_{max}$ which induce unphysical stationnary waves in $[0,r_{max}]$ in place of scattering states. The optical potential restores the physical meaning of the waves and makes non-hermitian the Hamiltonian. The potentials and the dipole moment are drawn figure \ref{potentials}.
The Hilbert space $\mathcal H$ is infinite dimensional, it is needed to represent it on a finite basis. We consider a discretisation $\{r_i = (i-1) \frac{r_{max}}{N} \}_{i=1,...,N_{DVR}}$ of $[0,r_{max}]$, and we consider the DVR basis (Discrete Variable Representation) $(\zeta_i(r))_i$ associated with the collocation points:
\begin{equation}
\zeta_i(r) = \frac{1}{\sqrt{N_{DVR} r_{max}}} \sum_{j=1}^{N_{DVR}} e^{\imath k_j (r-r_i)}
\end{equation}
with $k_j = \frac{2\pi (j-1-\frac{N_{DVR}}{2})}{r_{max}}$. We have chose $N_{DVR}=100$ and then $\dim \mathcal H^{DVR} = 200$ where $\mathcal H^{DVR}$ is the Hilbert space used in the numerical representation of the system and for which the DVR basis $(\zeta_i)_i$ times the electronic basis $(|{^2}\Sigma^+_g \rangle,{^2}\Sigma^+_u \rangle)$ constitutes the canonical basis. For off-field, the spectrum of the Hamiltonian is represented figure \ref{spectrum}.
We denote by $\{|gi\rangle\}_i$ the bound states for $W=0$ and by $\{\phi_{gi}(\vec R)\}_i$ the states issued from $\{|gi\rangle\}_i$ for $W > 0$. Following the works \cite{jaouadi, leclerc}, $\lambda_{g8}(\vec R)$ and $\lambda_{g9}(\vec R)$ the eigenvalues of $H(\vec R)$ issued from the 8th and the 9th bound states, have an exceptional point of coalescence at $\vec R_*=(W_*=0.197\ a.u.; \omega_*= 0.10242\ a.u.)$ ($3.948 \times 10^{13}\ W.cm^{-2}$ and $444.92\ nm$). Let $\mathcal C$ be the path surrounding two times $\vec R_*$ defined by
\begin{eqnarray}
W(t) & = & W_0 (1-\cos(4 \pi \frac{t}{T})) \\
\omega(t) & = & \omega_* + \Delta \omega \sin(4\pi \frac{t}{T})
\end{eqnarray}
with $W_0 = 0.105\ a.u.$ and $\Delta \omega = 0.0005\ a.u.$. The path is drawn figure \ref{path}.
As shown in  \cite{jaouadi, leclerc} the system is very weakly adiabatic, we have then chosen a very long interaction duration $T=40000\ a.u.$ in order to have a semblance of adiabatic behaviour. Figure \ref{eigenvalH2p} shows the evolution of the eigenvalues implicated in the exceptional crossing when this path is followed.

\subsubsection{Numerical integrations of the dynamics:}
\begin{figure}
\begin{center}
\includegraphics[width=10cm]{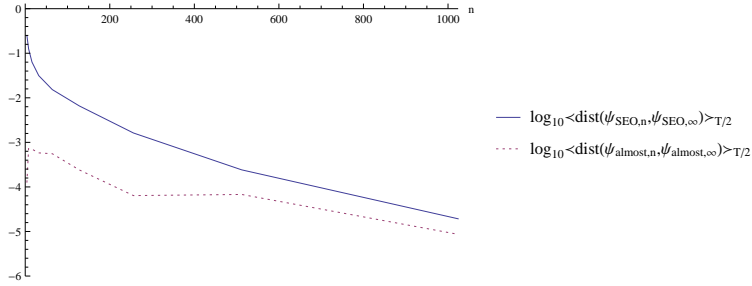}
\caption{\label{convergH2} Convergences of the wave packet propagation by time splitting of the evolution operator (SEO), and of the almost adiabatic representation in a logarithmic scale with respect to $n$ the number of time steps using in the propagation schemes during $[0,T]$. The convergence accuracies are computed with respect to reference solutions denoted $\psi_{*,\infty}$ computed with $2000$ time steps. The convergence accuracy for $n$ time steps is computed as being $ \prec \mathrm{dist}(\psi_{*,n},\psi_{*,\infty}) \succ_{T/2} = \frac{2}{T} \int_0^{T/2} \mathrm{dist}(\psi_{*,n}(t),\psi_{*,\infty}(t)) dt$ with $\mathrm{dist}(\psi,\phi) = 1 - \frac{|\langle \psi|\phi\rangle|}{\|\psi\| \|\phi\|}$.}
\end{center}
\end{figure}

We start with $\psi(0) = \phi_{g8}(0) = |g8 \rangle$ (which becomes the less dissipative state for small non zero $W$), and we consider the time-dependent wave function $\psi(t)$ solution of the Schr\"odinger equation $\ihbar \dot \psi = H(W(t),\omega(t)) \psi(t)$. We want to compare the adiabatic approximation and the almost adiabatic representation with a reference solution considered as the ``numerically exact'' solution. The second order differencing scheme cannot be used to obtain this reference solution, since for the time steps considered here, this scheme strongly diverges (the norm of the wave function becomes very large ($> 10^{100}$) rather than decreasing). To compute the reference solution, we consider the wave packet propagation based on the time splitted evolution operator:
\begin{equation}
U(T,0) = U(n\Delta t,(n-1)\Delta t) U((n-1)\Delta t,(n-2)\Delta t)...U(\Delta t,0)
\end{equation}
For sufficiently small time step $\Delta t$, $H(W(t),\omega(t))$ does not change significantly during $\Delta t$ (a such assumption is in accordance with the (almost) adiabatic assumption) and we can use the approximation
$$ U((k+1)\Delta t,k\Delta t) \simeq e^{-\ihbar^{-1} H(W(k \Delta t), \omega(k \Delta t)) \Delta t} $$
Finally the matrix exponentials $e^{-\ihbar^{-1} H(W(k \Delta t), \omega(k \Delta t)) \Delta t} $ are computed by a diagonalization of $H$ at each time step.\\
For the adiabatic approximation we have
\begin{equation}
\psi_{adiab}(t) = e^{-\ihbar^{-1}\int_0^t \lambda_{g8}(t')dt'} \phi_{g8}(t) 
\end{equation}
where $\phi_{g8}(t)$ must be defined by continuity with respect to $t$ taking into account the changes of Riemann sheet (which are located at $T/5$ and $7T/10$). We note that in this example, the geometric phase generators are still zero. Indeed $\langle \phi_{gi}*| = \langle \overline{\phi_{gi}}|$ (where $*$ denotes the biorthonormal vector and the overline denotes the complexe conjugation), we have then $\langle \phi_{gi}*|\dot \phi_{gi} \rangle = \int_0^{+\infty} \phi_{gi}(r,t) \dot \phi_{gi}(r,t) dr = \frac{1}{2} \frac{d}{dt} \int_0^{+\infty} \phi_{gi}(r,t)^2 dr = \frac{1}{2} \frac{d}{dt} \langle \phi_{gi}*| \phi_{gi} \rangle = 0$.\\
The almost adiabatic representation is
\begin{equation}
\psi_{almost}(t) = e^{-\ihbar^{-1}\int_0^t \lambda_{g8}(t')dt'-\int_0^t \eta(t')dt' } (\phi_{g8}(t)+x(t) \phi_{g9}(t))
\end{equation}
where we consider that $\phi_{g9}$ (the other state implicated in the exceptional point) is the only one state ruining the strict adiabaticity. In other words, we make the following assumption concerning the reduced wave operator:
\begin{equation}
X = x |\phi_{g9} \rangle \langle \phi_{g8}*| + \sum_{i\not=8,9} \epsilon_i |\phi_{gi} \rangle \langle \phi_{g8}*| \qquad\text{with } \epsilon_i \simeq 0
\end{equation}
This assumption consists then to consider that the strict adiabaticity is not valid for $\phi_{g9}$ but is valid for the other bound states. The equation \ref{eqY} is then reduced to
\begin{equation}
\label{eqx2}
\dot x(t) = \mathcal A_{g8,g9}(t) x(t)^2 - \ihbar^{-1}(\lambda_{g9}(t) - \lambda_{g8}(t)) x(t) + \mathcal A_{g9,g8}(t)
\end{equation}
with $\mathcal A_{g8,g9} = \langle \phi_{g8}*|\dot \phi_{g9} \rangle$. We integrate equation \ref{eqx2} by using the Runge-Kutta (RK4) method. We note that the non-adiabatic couplings must be computed by using the formula
\begin{equation}
\mathcal A_{g8,g9} = \frac{\langle \phi_{g8}*| \left(\frac{\partial H}{\partial W} \dot W + \frac{\partial H}{\partial \omega} \dot \omega \right)|\phi_{g9} \rangle}{\lambda_{g9} - \lambda_{g8}}
\end{equation}
obtained by projecting onto $\langle \phi_{g8}*|$ the time derivation of $H\phi_{g9} = \lambda_{g9} \phi_{g9}$. The non-adiabatic couplings do not be evaluated by a finite difference method applied to $\langle \phi_{g8}*|\dot \phi_{g9} \rangle$ because it induces divergences of the RK4 algorithm.\\
We diagonalize $H(W(t),\omega(t))$ at each time step. This constitutes a common precomputation for all representations. We note that to compute $e^{-\ihbar^{-1} H(W(k \Delta t), \omega(k \Delta t)) \Delta t} $ we could also use a split operator method \cite{leforestier} consisting to split the operator between a potential part which is diagonal in the DVR basis and a kinematic part which is diagonal in a FBR basis (Finite Basis Representation) obtained by Fourier transformations of the DVR basis. The computation of the matrix exponential needs then basis changes between the DVR and the FBR basis. To compute the adiabatic and the almost adiabatic representations, we do not need the complete diagonalization of $H$. We need only $\phi_{g8}$ at each time step for the adiabatic representation and only $\phi_{g8}$ and $\phi_{g9}$ at each time step for the almost adiabatic representation. We could then use a partial diagonalization algorithm (as for example the RDWA algorithm \cite{jolicard}) based on a recursive procedure (we compute $\phi_{g8}(k \Delta t)$ by successive improvements starting from the test function $\phi_{g8}((k-1)\Delta t)$). The goal of this section is to compare the almost adiabatic representation to a wave packet propagation method. We have then chosen a common diagonalization procedure for the two computations to the comparison be independent of it.\\
The convergences of the wave packet propagation and of the almost adiabatic representation are shown figure \ref{convergH2}.
We remark that the almost adiabatic representation is better converged than the reference solution. This do not mean that it is a better solution of the Schr\"odinger equation. It is natural that the convergence of the almost adiabatic representation be more easy since it consists to integrate a Schr\"odinger equation in a (time dependent) two dimensional space generated by $\{\phi_{g8},\phi_{g9}\}$ whereas the wave packet propragation consists to integrate a Schr\"odinger equation in a (time independent) 200 dimensional space. Nevertheless the accuracy of the almost adiabatic representation depends also on the validity of the assumption consisting to ignore the other states.\\
Table \ref{speed} shows the durations of the different computations.
\begin{table}
\begin{center}
\caption{\label{speed} CPU times of the computations of the wave packet propagation by time splitting of the evolution operator (SEO), of the adiabatic representation (AR), of the almost adiabatic representation (AAR) and of the second order differencing scheme (SOD), with $n=800$ time steps in $[0,T]$. We have shown the CPU times for only the propagation without the duration of the precomputation (diagonalisations at each step) and the CPU times including the precomputation. SOD method is not converged but is presented here only for comparison (we can note that for $8000$ time steps, SOD is still not converged and its CPU time is approximately multiplied by $10$).}
\begin{tabular}{r|cc}
 & \multicolumn{2}{c}{CPU time (s)} \\
 & excluding & including \\
 & the time of & the time of \\
 & diagonalization & diagonalization \\
\hline
SEO & 49.095 & 1922.644 \\
AR & 0.008 & 1873.597 \\
AAR & 0.064 & 1873.665 \\
(not converged) SOD & 879.299 & \textit{not needed}
\end{tabular}
\end{center}
\end{table}
For the viewpoint of the propagation only (without including the diagonalization time) adiabatic and almost adiabatic representations are very faster than the wave packet propagation. In this example this concerns a small time ($49\ s$) but for more time steps and with a quantum system having more degrees of freedom the advantage could be more significant. We can remark that in the previous example, the almost adiabatic representation consisted to the replacement of a linear two dimensional equation by a non-linear one dimensional equation. Its interest could seem small since it is more difficult to use a non-linear equation. But in this example, the almost adiabatic representation consists to the replacement of a linear 200 dimensional equation by a non-linear \textit{one} dimensional equation. If the accuracy of the approximation is satisfactory, the interest seems more important. The following section treats this point.

\subsubsection{The dynamics and its representations:}
\begin{figure}
\begin{center}
\includegraphics[width=12cm]{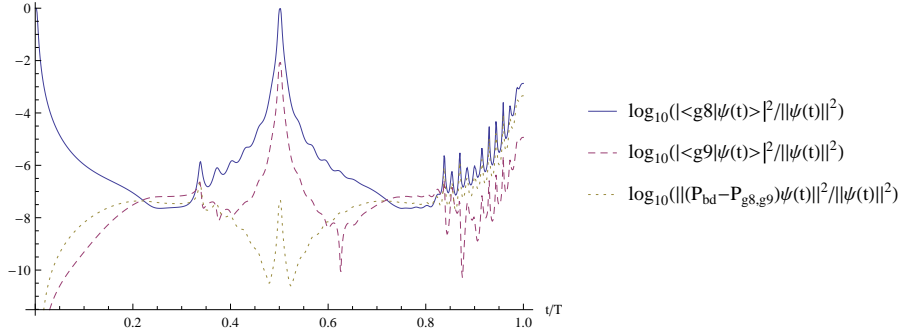}
\caption{\label{psiH2} Evolution of the populations of the states $|g8\rangle$, $|g9\rangle$, and of the total population of the other bound state. $P_{bd}$ denotes the projection onto the bound states and $P_{g8,g9}$ the projection onto the space spaned by $\{|g8\rangle,|g9\rangle\}$.}
\end{center}
\end{figure}

\begin{figure}
\begin{center}
\includegraphics[width=6cm]{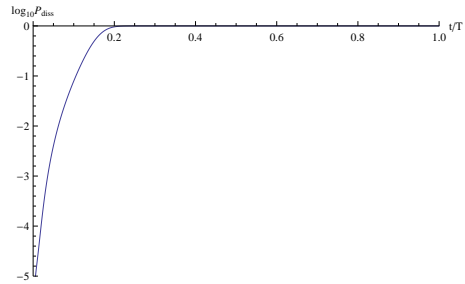}
\caption{\label{Pdiss} Evolution of the dissociation probability of the molecule.}
\end{center}
\end{figure}

\begin{figure}
\begin{center}
\includegraphics[width=12cm]{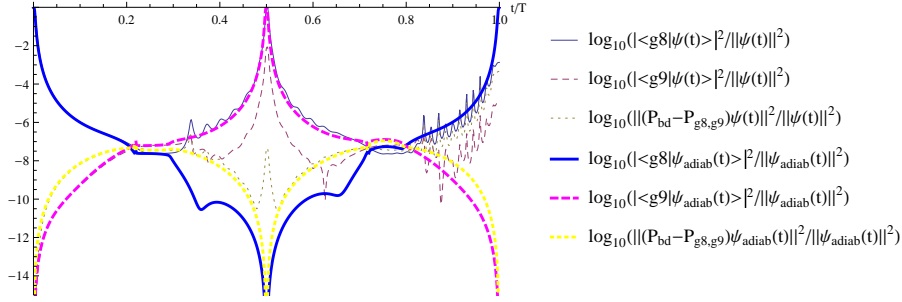}
\caption{\label{psiadiab} Evolution of the populations of the states $|g8\rangle$, $|g9\rangle$, and of the total population of the other bound states in the adiabatic approximation (thick lines). In order to facilitate the comparison, we have recall also the true evolution (thin lines).  $P_{bd}$ denotes the projection onto the bound states and $P_{g8,g9}$ the projection onto the space spaned by $\{|g8\rangle,|g9\rangle\}$.}
\end{center}
\end{figure}

\begin{figure}
\begin{center}
\includegraphics[width=12cm]{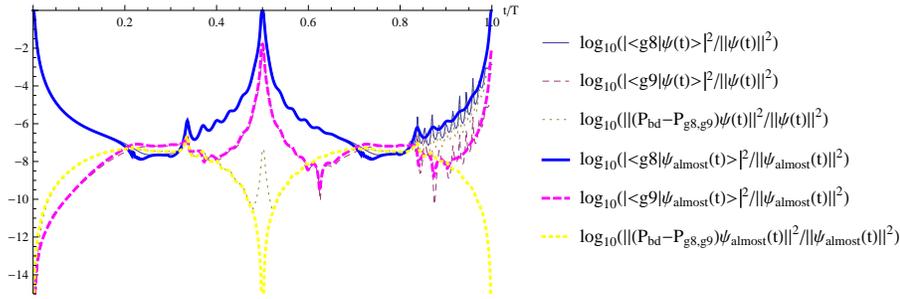}
\caption{\label{psiWO} Evolution of the populations of the states $|g8\rangle$, $|g9\rangle$, and of the total population of the other bound states in the almost adiabatic approximation (thick lines). In order to facilitate the comparison, we have recall also the true evolutions (thin lines).  $P_{bd}$ denotes the projection onto the bound states and $P_{g8,g9}$ the projection onto the space spaned by $\{|g8\rangle,|g9\rangle\}$.}
\end{center}
\end{figure}

\begin{figure}
\begin{center}
\includegraphics[width=10cm]{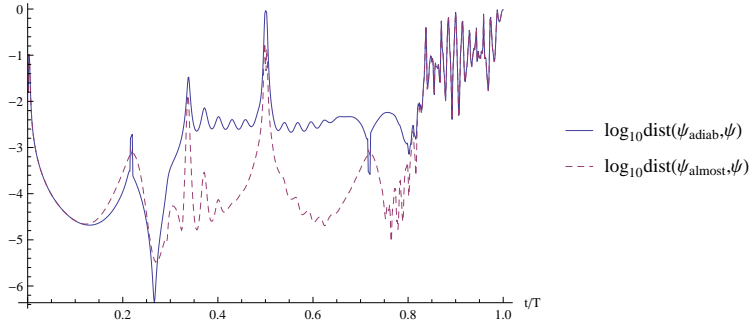}
\caption{\label{comperr} Evolution of the error concerning the evaluation of the wave function for the adiabatic and the almost adiabatic approximations. The distance is defined by $\mathrm{dist}(\phi,\phi) = 1 - \frac{|\langle \psi|\phi \rangle|}{\|\psi\|\|\phi\|}$. The strong error observed at the end of the second turn is due to the bound states not included in the active space of the almost adiabatic representation (see figure \ref{psiWO}).}
\end{center}
\end{figure}

The reference solution of the dynamics is shown figure \ref{psiH2}.
As explained in \cite{jaouadi, leclerc}, although the interaction duration $T$ is very long, the expected adiabatic state inversion between $|g8\rangle$ and $|g9\rangle$ after one turn does not occur. After two turns, the problem is still more important since other bound states are significantly occuped. It is ridiculous to try a larger interaction duration to have more adiabatic behaviours since with this interaction duration the molecule is already ``completely'' dissociated as we can see it figure \ref{Pdiss}.
Figure \ref{psiadiab} shows the adiabatic representation of the dynamics.
From the middle of the first turn, the adiabatic approximation fails completely to represent the dynamics (this corresponds to the moment where the followed eigenvector becomes the more dissipative, this failure of the adiabatic approximation is then coherent with the adiabatic theorems for non-self-adjoint Hamiltonians \cite{nenciu2,joye}). The almost adiabatic representation of the dynamics is shown figure \ref{psiWO}.
The almost adiabatic approximation reproduces with very good satisfaction the behaviours of the populations of $|g8\rangle$ and $|g9\rangle$. At the end of the second turn, only the population of the other bound states is not correctly reproduced. This is natural since we have ignored them in the active space of the wave operator. It could possible to correct this small problem by including one or two other bound states (spectraly close to $\{\phi_{g8},\phi_{g9}\}$) in the active space. The errors of the adiabatic approximation and of the almost adiabatic approximation are shown figure \ref{comperr}.

\section{Conclusion}
The role of the wave operators in quantum dynamics can be now resumed. The quantum dynamics can be represented by the projector $P(t)$ solution of $\ihbar \dot P = [H,P]$ with $P^2=P$. $P(t)$ is the projector onto the space spaned by all the solutions of the Schr\"odinger equation for initial conditions within $\Ran P(0)$. The non-linear analogue of $P$ is the time-dependent wave operator $\Omega(t)$ solution of $\ihbar \dot \Omega = [H,\Omega]\Omega$ with $\Omega^2=\Omega$. $\Omega(t)$ is the comparison between the true quantum dynamics and the effective dynamics within $\Ran P(0)$ if $\dist(P(t),P_0(0))< \frac{\pi}{2}$. At the adiabatic limit, $\dist(P(t),P_0(t))\sim 0$, the quantum dynamics can be approached within the active space $\Ran P_0(t)$ where $P_0(t)$ is an eigenprojector, $[H,P_0]=0$ and $P_0^2=P_0$. The non-linear analogue of $P_0(t)$ is the Bloch wave operator $\Omega(t)$ solution of $[H,\Omega]\Omega=0$ with $\Omega^2=\Omega$. $\Omega(t)$ is the comparison between the states from $\Ran P_0(t)$ and $\Ran P_0(0)$. Finally in the almost adiabatic situation, $\dist(P(t),P_0(t))< \frac{\pi}{2}$, the generalized time-dependent wave operator $\Omega(t)$ solution of $\ihbar \dot \Omega = [H,\Omega]\Omega + \ihbar \Omega \dot \Omega$ is a comparison between the true dynamics  within $\Ran P(t)$ and the effective dynamics within $\Ran P_0(t)$. For the three approximations, the effective Hamiltonian governing the approximate dynamics within the active space, can be written with the general formula $H^{eff} = P_0 H \Omega - \ihbar \dot P_0 \Omega$ (with for the adiabatic limit $\Omega = P_0(P_0P_0P_0)^{-1}=P_0$ and $\dot P_0 = 0$ for a fixed active space).\\
In practice, the use of the almost adiabatic representation can be efficient to treat problems where the adiabatic approximation fails as the dynamics surrounding exceptional points. By mixing adiabatic approximation and wave operator method, we can treat such a problem by a partial diagonalization of the time-dependent hamiltonian (to find the two eigenvectors implicated in the coalescence) and by a propagation of a wave operator in a one dimensional space (even if the dimension of the Hilbert space is larger than 2, the components of $X$ on the vectors which are not implicated by the coalescence can be neglected because they are governed by very small non-adiabatic coupling terms).

\appendix
\section{Demonstration of the Bloch equation for the generalized case}
\label{appA}
\subsection{The projector into the limbo space}
Let $U \in \mathcal U(\mathcal H)$ be the evolution operator:
\begin{equation}
\ihbar \dot U = H U
\end{equation}
Since $\ihbar \dot P = [H,P]$ with $P(0) = P_0(0)$ we have
\begin{equation}
P = UP_0(0)U^{-1}
\end{equation}
Let $K \in \mathcal B(\mathcal H)$ be the operator defined by
\begin{equation}
K = \ihbar (\dot P_0 P_0 + \dot Q_0 Q_0)
\end{equation}
where $Q_0 = 1-P_0$ is the projector into $\Ran P_0^\bot$. $K$ is the usual adiabatic kernel used in each demonstration of the adiabatic theorems \cite{messiah, nenciu,nenciu2,joye}. Let $V \in \mathcal U(\mathcal H)$ be the evolution operator associated with $K$:
\begin{equation}
\ihbar \dot V = KV
\end{equation}
A classical property of $K$ \cite{messiah,nenciu,nenciu2,joye} easily verifiable is that
\begin{equation}
P_0(t) = V(t)P_0(0)V(t)^{-1}
\end{equation}
It follows that
\begin{equation}
P = UV^{-1} P_0 VU^{-1}
\end{equation}

\subsection{Lemma : derivative of the inverse within $\Ran P_0(t)$}
Let $W \in \mathcal B(\mathcal H)$ be an arbitrary operator such that $(P_0 W P_0)^{-1}$ exists. By definition we have
\begin{equation}
(P_0WP_0)^{-1} P_0WP_0 = P_0
\end{equation}
By derivating this expression we find
\begin{equation}
\frac{d(P_0WP_0)^{-1}}{dt} P_0WP_0 + (P_0WP_0)^{-1} \frac{d(P_0WP_0)}{dt} = \dot P_0
\end{equation}
and then
\begin{eqnarray}
\frac{d(P_0WP_0)^{-1}}{dt} P_0 & = & \dot P_0 (P_0 W P_0)^{-1} \nonumber \\
& & \qquad - (P_0WP_0)^{-1} \frac{d(P_0WP_0)}{dt} (P_0WP_0)^{-1}
\end{eqnarray} 

\subsection{Derivative properties of the generalized time-dependent wave operator}
\begin{equation}
\Omega P_0 = \Omega \Rightarrow \dot \Omega = \dot \Omega P_0 + \Omega \dot P_0
\end{equation}

\begin{equation}
P_0 \Omega = P_0 \Rightarrow \dot P_0 \Omega + P_0 \dot \Omega = \dot P_0 \Rightarrow \Omega \dot \Omega = \Omega \dot P_0 - \Omega \dot P_0 \Omega
\end{equation}

\subsection{Proof of the Bloch equation}
\begin{eqnarray}
\Omega & = & P(P_0PP_0)^{-1} \\
& = & UV^{-1} P_0 VU^{-1} (P_0 UV^{-1} P_0 VU^{-1} P_0)^{-1} \\
& = & UV^{-1} P_0 VU^{-1} P_0 (P_0 VU^{-1} P_0)^{-1} (P_0 UV^{-1} P_0)^{-1} \\
& = & UV^{-1} (P_0 UV^{-1} P_0)^{-1}
\end{eqnarray}

By using the derivative properties of $\Omega$ and of $(P_0WP_0)^{-1}$ we have
\begin{eqnarray}
\ihbar \dot \Omega & = & \ihbar \dot \Omega P_0 + \ihbar \Omega \dot P_0 \\
& = & \ihbar \dot U V^{-1}(P_0 UV^{-1} P_0)^{-1} - \ihbar UV^{-1} \dot V V^{-1} (P_0UV^{-1})^{-1} \nonumber \\
& & + \ihbar UV^{-1} \dot P_0(P_0 UV^{-1}P_0)^{-1} \nonumber \\
& & - \ihbar UV^{-1} (P_0UV^{-1}P_0)^{-1} \left(\dot P_0UV^{-1}P_0+P_0 \dot UV^{-1} P_0 \right. \nonumber \\
& & \quad \quad \left. - P_0 UV^{-1}\dot VV^{-1}P_0+P_0UV^{-1}\dot P_0 \right)(P_0UV^{-1}P_0)^{-1} \nonumber \\
& & + \ihbar \Omega \dot P_0 \\
& = & H\Omega - UV^{-1}K(P_0UV^{-1}P_0)^{-1}+\ihbar UV^{-1} \dot P_0 (P_0UV^{-1}P_0)^{-1} \nonumber \\
& & - \ihbar \Omega \dot P_0 \Omega - \Omega H \Omega \nonumber \\
& & + \Omega UV^{-1}K(P_0UV^{-1}P_0)^{-1} - \ihbar \Omega UV^{-1} \dot P_0 (P_0 UV^{-1} P_0)^{-1} \nonumber \\
& & + \ihbar \Omega \dot P_0
\end{eqnarray}

By using the expression of $\Omega \dot \Omega$ and the fact that $KP_0 = \ihbar \dot P_0 P_0$ we find
\begin{equation}
\ihbar \dot \Omega = H\Omega - \Omega H \Omega + \ihbar \Omega \dot \Omega
\end{equation}

\subsection{Effective Hamiltonian}
\label{appA5}
Let $U^{eff} \in \mathcal L(\Ran P_0(0),\Ran P_0(t))$ be such that
\begin{equation}
U P_0(0) = \Omega U^{eff}
\end{equation}
$U \in \mathcal U(\mathcal H)$ being the evolution operator. By derivating this expression we find
\begin{eqnarray}
\dot U P_0(0) & = & \dot \Omega U^{eff} + \Omega \dot U^{eff} \\
HUP_0(0) & = & H\Omega U^{eff} - \Omega H \Omega U^{eff} + \ihbar \Omega \dot \Omega U^{eff} + \ihbar \Omega \dot U^{eff}
\end{eqnarray}
By projecting this equation on $P_0$ we find
\begin{equation}
P_0HUP_0(0) = P_0 H \Omega U^{eff} - P_0 H \Omega U^{eff} + \ihbar P_0 \dot \Omega U^{eff} + \ihbar P_0 \dot U^{eff}
\end{equation}
Since by definition $UP_0(0) = \Omega U^{eff}$ and $P_0 \dot \Omega = \dot P_0 - \dot P_0 \Omega$, we have
\begin{equation}
P_0 H \Omega U^{eff} = \ihbar \dot P_0 U^{eff} - \ihbar \dot P_0 \Omega U^{eff} + \ihbar P_0 \dot U^{eff}
\end{equation}
Finally since $P_0 U^{eff} = U^{eff}$ we have $\dot P_0 U^{eff} + P_0 \dot U^{eff} = \dot U^{eff}$ and then
\begin{equation}
P_0H\Omega U^{eff} = \ihbar \dot U^{eff} - \ihbar \dot P_0 \Omega U^{eff}
\end{equation}
\begin{equation}
\ihbar \dot U^{eff} = (P_0H- \ihbar \dot P_0) \Omega U^{eff}
\end{equation}
We can then write $H^{eff} = (P_0H - \ihbar \dot P_0) \Omega$.

\section{Proof of the equation of the modified reduced wave operator}
\label{appB}
Let $X=Q_0\Omega P_0$ and $H_{adiab} = H - K$ (with $K = \ihbar (\dot P_0P_0 + \dot Q_0 Q_0)$ and $Q_0+P_0=1$).
\subsection{Expression of $(Q_0-X)H_{adiab}(P_0+X)$}
\begin{eqnarray}
& & (Q_0-X)H_{adiab}(P_0+X) \nonumber \\
& &=  (Q_0-X)H(P_0+X) - \ihbar (Q_0-X)\dot P_0P_0 - \ihbar (Q_0-X)\dot Q_0 X \\
& &=  (Q_0-X)H(P_0+X) - \ihbar (Q_0-X)\dot P_0P_0 + \ihbar (Q_0-X)\dot P_0 X \\
& & =  (Q_0-X)H(P_0+X) - \ihbar (Q_0-X)\dot P_0(P_0-X)
\end{eqnarray}
Since $(Q_0-X)P_0 = - X$ we have by using $XQ_0 = 0 \Rightarrow \dot XQ_0 = - X \dot Q_0$:
\begin{eqnarray}
(Q_0-X)\dot P_0 & = & -\dot X - (\dot Q_0 - \dot X) P_0 \\
& = & \dot X (P_0-1) - \dot Q_0 P_0 \\
& = & -  \dot X Q_0 + \dot P_0 P_0 \\
& = & X \dot Q_0 + \dot P_0 P_0 \\
& = & -X \dot P_0 + \dot P_0 P_0
\end{eqnarray}
Moreover $XP_0 = X \Rightarrow \dot X P_0+X \dot P_0 = \dot X \Rightarrow X \dot P_0 = \dot X(1-P_0) \Rightarrow X\dot P_0 P_0 = 0$. We have then
\begin{equation}
(Q_0-X)H_{adiab}(P_0+X) = (Q_0-X)H(P_0+X) - \ihbar \dot P_0 P_0  - \ihbar X \dot P_0 X
\end{equation}

\subsection{Equation of the reduced wave operator}
By the equation $\ihbar \dot \Omega = H \Omega - \Omega H \Omega + \ihbar \Omega \dot \Omega$ we find (by using $P_0 \dot \Omega = \dot P_0 - \dot P_0 \Omega$)
\begin{eqnarray} 
\ihbar \dot X & = & H \Omega - (P_0+X)H\Omega + \ihbar(P_0+X)\dot \Omega - \ihbar \dot P_0 \\
& = & (Q_0-X)H\Omega + \ihbar P_0 \dot \Omega + \ihbar X \dot \Omega - \ihbar \dot P_0 \\
& = & (Q_0-X)H\Omega + \ihbar \dot P_0 - \ihbar \dot P_0 \Omega + \ihbar X(\dot P_0 - \dot P_0\Omega) - \ihbar \dot P_0 \\
& = & (Q_0-X)H\Omega -\ihbar \dot P_0 \Omega + \ihbar X \dot P_0 - \ihbar X \dot P_0 \Omega \\
& = & (Q_0-X)H\Omega - \ihbar (1+X) \dot P_0 \Omega + \ihbar X \dot P_0 \\
& = & (Q_0-X)H\Omega - \ihbar (1+X) \dot P_0 P_0 - \ihbar (1+X) \dot P_0 X \nonumber \\
& & \qquad + \ihbar X \dot P_0 \\
& = & (Q_0-X)H\Omega - \ihbar (1+X) \dot P_0 P_0 - \ihbar \dot P_0 X - \ihbar X \dot P_0 X \nonumber \\
& & \qquad + \ihbar X \dot P_0 \\
& = & (Q_0-X)H\Omega - \ihbar \dot P_0 P_0 - \ihbar \dot P_0 X - \ihbar X \dot P_0 X + \ihbar X \dot P_0 \\
& = & (Q_0-X) H_{adiab} (P_0+X) - \ihbar[\dot P_0,X]
\end{eqnarray}
where we have used $X \dot P_0 P_0 = 0$.

\subsection{Passage to the modified reduced wave operator}
Since $P_0^2 = P_0$, $Q_0^2 = Q_0$ and $P_0+Q_0=1$ we have $\dot P_0 P_0 + \dot Q_0 Q_0 = - P_0 \dot P_0 - Q_0 \dot Q_0$.
\begin{eqnarray}
[K,X] & = & \ihbar(\dot P_0 P_0 X + \dot Q_0 Q_0 X + X P_0 \dot P_0 + X Q_0 \dot Q_0 ) \\
& = & \ihbar \dot Q_0 X + \ihbar X \dot P_0 \\
& = & - \ihbar \dot P_0 X + \ihbar X \dot P_0 \\
& = & -\ihbar[\dot P_0,X]
\end{eqnarray}
We have then
\begin{equation}
\ihbar \dot X -[K,X] = (Q_0-X) H_{adiab} (P_0+X)
\end{equation}
Let $V \in \mathcal U(\mathcal H)$ be such that $\ihbar \dot V = KV$ and $Y = V^{-1}XV$.
\begin{eqnarray}
\ihbar \dot X & = & \ihbar \dot VYV^{-1} + \ihbar V \dot Y V^{-1} - \ihbar VYV^{-1} \dot VV^{-1} \\
& = & KX + \ihbar V\dot YV^{-1} - XK \\
& = & [K,X] + \ihbar V\dot YV^{-1}
\end{eqnarray}
We have then
\begin{equation}
\ihbar \dot Y = V^{-1} (Q_0-X) H_{adiab} (P_0+X) V
\end{equation}
Since $V(t)P_0(0) = P_0(t) V(t)$ we have
\begin{equation}
\ihbar \dot Y = (Q_0(0)-Y)V^{-1}H_{adiab}V(P_0(0)+Y)
\end{equation}

\subsection{The intertwining operator}
\label{appB4}
Let $\{\phi_a(t)\}_{a=1,...,m}$ be an orthonormal basis of $\Ran P_0(t)$ and $\{\phi_a(t)\}_{a>m}$ be an orthonormal basis of $\Ran P_0(t)^\bot$. Let $P_a = |\phi_a \rangle \langle \phi_a|$, we have
\begin{equation}
K = \ihbar (\dot P_0 P_0 + \dot Q_0 Q_0) = \ihbar \left( \sum_{a,b=1}^m \dot P_a P_b + \sum_{a,b > m} \dot P_a P_b \right) 
\end{equation}
$V(t) P_0(0) = P_0(t) V(t)$ and $V(t) Q_0(0) = Q_0(t) V(t)$ imply that
\begin{equation}
V(t) \phi_a(0) = \left\{\begin{array}{cc} \sum_{b=1}^{m} \langle \phi_b(t)|V(t)|\phi_a(0)\rangle \phi_b(t) & \text{if } a\leq m \\ \sum_{b>m} \langle \phi_b(t)|V(t)|\phi_a(0)\rangle \phi_b(t) & \text{if } a> m \end{array} \right.
\end{equation}
and then
\begin{equation}
V(t) = \sum_{a,b=1}^m V_{ba}(t) |\phi_b(t) \rangle \langle \phi_a(0)| + \sum_{a,b>m} V_{ba}(t) |\phi_b(t) \rangle \langle \phi_a(0)|
\end{equation}
By injecting this expression in $\ihbar \dot V = KV$ and by projecting on the left on $\langle \phi_b(t)|$ and on the right on $|\phi_a(0)\rangle$ we find
\begin{equation}
\dot V_{ba}(t) = - \left\{\begin{array}{cc} \sum_{c=1}^m \langle \phi_b(t) | \dot \phi_c(t) \rangle V_{ca}(t) & \text{if } a \leq m \\ \sum_{c>m} \langle \phi_b(t) | \dot \phi_c(t) \rangle V_{ca}(t) & \text{if } a > m \end{array} \right.
\end{equation}
Finally we have
\begin{equation}
V(t) = \sum_{a,b} \left[ \Te^{- \int_0^t \mathfrak A(t')dt'} \right]_{ba} |\phi_b(t) \rangle \langle \phi_a(0) |
\end{equation}
with
\begin{equation}
\mathfrak A(t) = \left( \begin{array}{cc}
\mathfrak A_{11} & 0 \\ 0 & \mathfrak A_{22}
\end{array} \right)
\end{equation}
with
\begin{equation}
\mathfrak A_{11} =  \left( \begin{array}{ccc}
\langle \phi_1(t)|\dot \phi_1(t) \rangle & ... & \langle \phi_1(t)|\dot \phi_m(t) \rangle \\
\vdots & \ddots & \vdots \\
\langle \phi_m(t)|\dot \phi_1(t)  \rangle & ... & \langle \phi_m(t)|\dot \phi_m(t)  \rangle \\
\end{array} \right)
\end{equation}
and
\begin{equation}
\mathfrak A_{22} =  \left( \begin{array}{ccc}
 \langle \phi_{m+1}(t)|\dot \phi_{m+1}(t)  \rangle & ... & \langle \phi_{m+1}(t)|\dot \phi_{\dim \mathcal H}(t) \rangle \\
 \vdots & \ddots & \vdots \\
 \langle \phi_{\dim \mathcal H}(t)|\dot \phi_{m+1}(t) \rangle & ... & \langle \phi_{\dim \mathcal H}(t)|\dot \phi_{\dim \mathcal H}(t) \rangle
\end{array} \right)
\end{equation}

\section{Splitted second order differencing scheme}
\label{appC}
Let $H(t) = H_0(t) + D$ with $H_0(t)^\dagger = H_0(t)$ and $D^\dagger = - D^\dagger$. $H_0$ is the ``energy'' part of the Hamiltonian and $D$ is the dissipative part. In the example treated section V, we have
\begin{equation}
H_0 = \frac{\hbar}{2}\left(\begin{array}{cc} 0 & \Omega \\ \Omega & 2 \Delta \end{array} \right) \qquad D = \left(\begin{array}{cc} 0 & 0 \\ 0 & - \imath \frac{\Gamma}{4} \end{array} \right)
\end{equation}
Let $U(t+\Delta t,t)$ be the evolution operator associated with $H(t)$ and $U_0(t+\Delta t,t)$ be the evolution operator associated with $H_0(t)$.
\begin{eqnarray}
\psi(t+\Delta t) & = & U(t+\Delta t ,t) \psi(t) \\
& = & e^{-\ihbar^{-1} D \Delta t + \mathcal O(\Delta t^2)} U_0(t+\Delta t,t) \psi(t) \\
& = & e^{-\ihbar^{-1} D \Delta t + \mathcal O(\Delta t^2)} \psi_0(t+\Delta t)
\end{eqnarray}
where $\psi_0(t+\Delta t)$ is solution of
\begin{equation}
\forall s \in [t,t+\Delta t], \quad \ihbar \frac{d}{ds} \psi_0(s) = H_0(s) \psi_0(s) \qquad \psi_0(t) = \psi(t)
\end{equation}
We have then
\begin{eqnarray}
\psi_0(t+\Delta t) & = & \psi_0(t) + \dot \psi_0(t) \Delta t + \ddot \psi_0(t) \frac{\Delta t^2}{2} + \mathcal O(\Delta t^3) \\
& = & \psi(t) -\ihbar^{-1} H_0(t) \psi(t) \Delta t + \ddot \psi_0(t) \frac{\Delta t^2}{2} + \mathcal O(\Delta t^3)
\end{eqnarray}
and
\begin{eqnarray}
\psi(t+\Delta t) & = & e^{-\ihbar^{-1} D \Delta t + \mathcal O(\Delta t^2)} \nonumber \\
& & \times \left(\psi(t) -\ihbar^{-1} H_0(t) \psi(t) \Delta t + \ddot \psi_0(t) \frac{\Delta t^2}{2} + \mathcal O(\Delta t^3) \right)
\end{eqnarray}
In a same way, we have
\begin{eqnarray}
\psi(t-\Delta t) & = & e^{\ihbar^{-1} D \Delta t + \mathcal O(\Delta t^2)} \nonumber \\
& & \times \left(\psi(t) +\ihbar^{-1} H_0(t) \psi(t) \Delta t + \ddot \psi_0(t) \frac{\Delta t^2}{2} + \mathcal O(\Delta t^3) \right)
\end{eqnarray}
Finally
\begin{eqnarray}
& & \psi(t+\Delta t) - e^{- 2\ihbar^{-1} D \Delta t + \mathcal O(\Delta t^2)} \psi(t-\Delta t) \nonumber \\
& &= e^{-\ihbar^{-1} D \Delta t + \mathcal O(\Delta t^2)} \left(-2 \ihbar^{-1} H_0(t) \psi(t) \Delta t + \mathcal O(\Delta t^3) \right)
\end{eqnarray}
We have then the following propagation scheme for a partition $\{t_0,...,t_N\}$ of $[0,T]$ ($t_{i+1}-t_i = \Delta t$):
\begin{equation}
\psi_{n+1} = e^{- 2\ihbar^{-1} D \Delta t} \psi_{n-1} - 2 \ihbar^{-1} e^{-\ihbar^{-1} D \Delta t} H_0(t_n) \psi_n
\end{equation}

\end{document}